\documentclass{aa}
\usepackage{psfig}
\usepackage{natbib}
\usepackage{longtable}
\usepackage{graphicx}
\bibpunct{(}{)}{,}{a}{}{,} % to follow the A&A style

\def\la{\;
\raise0.3ex\hbox{$<$\kern-0.75em\raise-1.1ex\hbox{$\sim$}}\; }
\def\ga{\;
\raise0.3ex\hbox{$>$\kern-0.75em\raise-1.1ex\hbox{$\sim$}}\; }

\newcommand{\kms}{km~s$^{-1}$}
\newcommand{\kmspc}{km~s$^{-1}$~pc$^{-1}$}

\newcommand{\cm}{cm$^{-2}$}
\newcommand{\cmm}{cm$^{-3}$}
\newcommand{\etal}{{et al.}}
\newcommand{\nhhh}{NH$_3$}
\newcommand{\nnhp}{N$_2$H$^+$}
\newcommand{\Dv}{$\Delta v$}
\newcommand{\Tmb}{$T_{\scriptscriptstyle \rm MB}$}
\newcommand{\VLSR}{$V_{\scriptscriptstyle \rm LSR}$}
\newcommand{\Tkin}{$T_{\rm kin}$}

\newcommand{\Tex}{$T_{\rm ex}$}
\newcommand{\ta}{$\tau_{\scriptscriptstyle \rm 11}$}
\newcommand{\tb}{$\tau_{\scriptscriptstyle \rm 22}$}
\newcommand{\nHH}{$n_{\scriptscriptstyle {\rm H}_2}$}
\newcommand{\rds}{rad~s$^{-1}$}

\begin{document}

\title{A rotating helical filament in the L1251 dark cloud
}
\author{
S. A. Levshakov\inst{1,2,3}
\and
D. Reimers\inst{1}
\and
C. Henkel\inst{4,5}
}
\institute{
Hamburger Sternwarte, Universit\"at Hamburg,
Gojenbergsweg 112, D-21029 Hamburg, Germany
\and
Ioffe Physical-Technical Institute,
Polytekhnicheskaya Str. 26, 194021 St.~Petersburg, Russia
\and
St.~Petersburg Electrotechnical University `LETI', Prof. Popov Str. 5,
197376 St.~Petersburg, Russia \\
\email{lev@astro.ioffe.rssi.ru}
\and
Max-Planck-Institut f\"ur Radioastronomie, Auf dem H\"ugel 69, D-53121 Bonn, Germany
\and
Astronomy Department, King Abdulaziz University, P.O.
Box 80203, Jeddah 21589, Saudi Arabia
}
\date{Received 00  ; Accepted 00}
\abstract
{}
{We derive the physical properties of a filament discovered 
in the dark cometary-shaped cloud \object{L1251}.
}
{Mapping observations in the \nhhh(1,1) and (2,2) inversion lines,
encompassing 300 positions toward \object{L1251}, were performed with the
Effelsberg 100-m telescope at a spatial resolution of 40\arcsec\ and 
a spectral resolution of 0.045 \kms.
}
{The filament \object{L1251A} consists of three
condensations ($\alpha$, $\beta$, and $\gamma$) of elongated morphology,
which are combined in a long and narrow structure
covering a $38\arcmin \times 3\arcmin$ angular range ($\sim 3.3$ pc $\times\ 0.3$ pc).
Comparing the kinematics with the more extended envelope ($\sim 61\arcmin \times 33\arcmin$) 
emitting in $^{13}$CO,  we find that: 
$(1)$ the angular velocity of the envelope around the horizontal axis E$\to$W
is $\Omega^{\scriptscriptstyle\rm EW}_{\scriptscriptstyle {\rm CO}} \approx -2\times10^{-14}$ \rds\ 
(the line-of-sight velocity is more negative to the north); 
$(2)$ approximately one half of the filament (combined $\alpha$ and $\beta$ condensations)
exhibits counter-rotation with
$\Omega^{\scriptscriptstyle\rm EW}_{\alpha\beta} \approx 2\times10^{-14}$ \rds;
$(3)$ one third of the
filament (the $\gamma$ condensation) co-rotates with 
$\Omega^{\scriptscriptstyle\rm EW}_\gamma \approx -2\times10^{-14}$ \rds;
$(4)$ the central part of the filament between these two kinematically
distinct regions does not show any rotation around this axis;  
$(5)$ the whole filament revolves slowly around the vertical axis S$\to$N
with $\Omega^{\scriptscriptstyle\rm SN}_{\rm tot} \approx 7\times10^{-15}$ \rds.
The opposite chirality (dextral and sinistral) of the $\alpha\beta$ and $\gamma$ condensations
indicates magnetic field helicities of two types, negative and positive,
which were most probably caused by dynamo mechanisms.
We estimated the magnetic Reynolds number $R_{\rm m} \ga 600$ and the Rossby number ${\cal R} < 1$, which means
that dynamo action is important.
}
{}

\authorrunning{S. A. Levshakov \etal\ }

\titlerunning{A rotating helical filament in the L1251 dark cloud}

\maketitle

\section{Introduction}
\label{sect-1}

Observations show that magnetic fields are ubiquitous in the local universe
and are often associated with filamentary structures
(e.g., Andr\'e \etal\ 2014, and references cited therein).
Filamentary molecular clouds play an important role in star formation
since star-forming cores appear primarily along dense elongated blocks.
Thus to derive their physical parameters
is a crucial issue for understanding star formation (e.g., Konyves \etal\ 2015).
Magnetic fields support and form the morphology of filaments, and
there is evidence that some filaments are
wrapped by helical fields (Matthews \etal\ 2001;
Hily-Blant \etal\ 2004; Carlqvist \etal\ 2003; Poidevin \etal\ 2010).

In the present paper we describe a filament detected within the molecular cloud \object{L1251}
by means of spectral observations of \nhhh\ emission lines.
\object{L1251} is a star-forming dark cloud
of the opacity class 5 (the second highest, see Lynds 1962),
elongated E-W inside the molecular ring in the Cepheus flare (Hubble 1934; Lebrun 1986)
at a distance of $D = 300\pm50$ pc (Kun \& Prusti 1993).
With the coordinates $\ell \approx 115$\degr, $b \approx 15$\degr,
its distance from the galactic midplane is about 100 pc.
$^{13}$CO observations show a cometary distribution of gas
with a {\bf U}-shaped dense ``head'' turned toward the center
of the Cep~--~Cas void created by a $4\times10^4$ yr-old Type~I supernova
(McCammon \etal\, 1983; Grenier \etal\, 1989). 
Supernova shock fronts have probably triggered the formation of stars 
in \object{L1215} and affected the cloud morphology.

The detailed distribution of molecular gas in \object{L1251}
was studied by Grenier \etal\, (1989), Sato \& Fukui (1989), and
later on at higher angular resolutions by Sato \etal\, (1994, hereafter S94). 
S94 revealed five C$^{18}$O dense cores embedded in the $^{13}$CO cloud. 
The cores, designated as ``A'' to ``E'' with increasing R.A.,
exhibit an elongated structure with a major axis of $\sim 5$ pc (E$\to$W)
and a minor axis of $\sim 3$ pc (S$\to$N).

\begin{figure}[t]
\vspace{0.0cm}
\hspace{-1.5cm}\psfig{figure=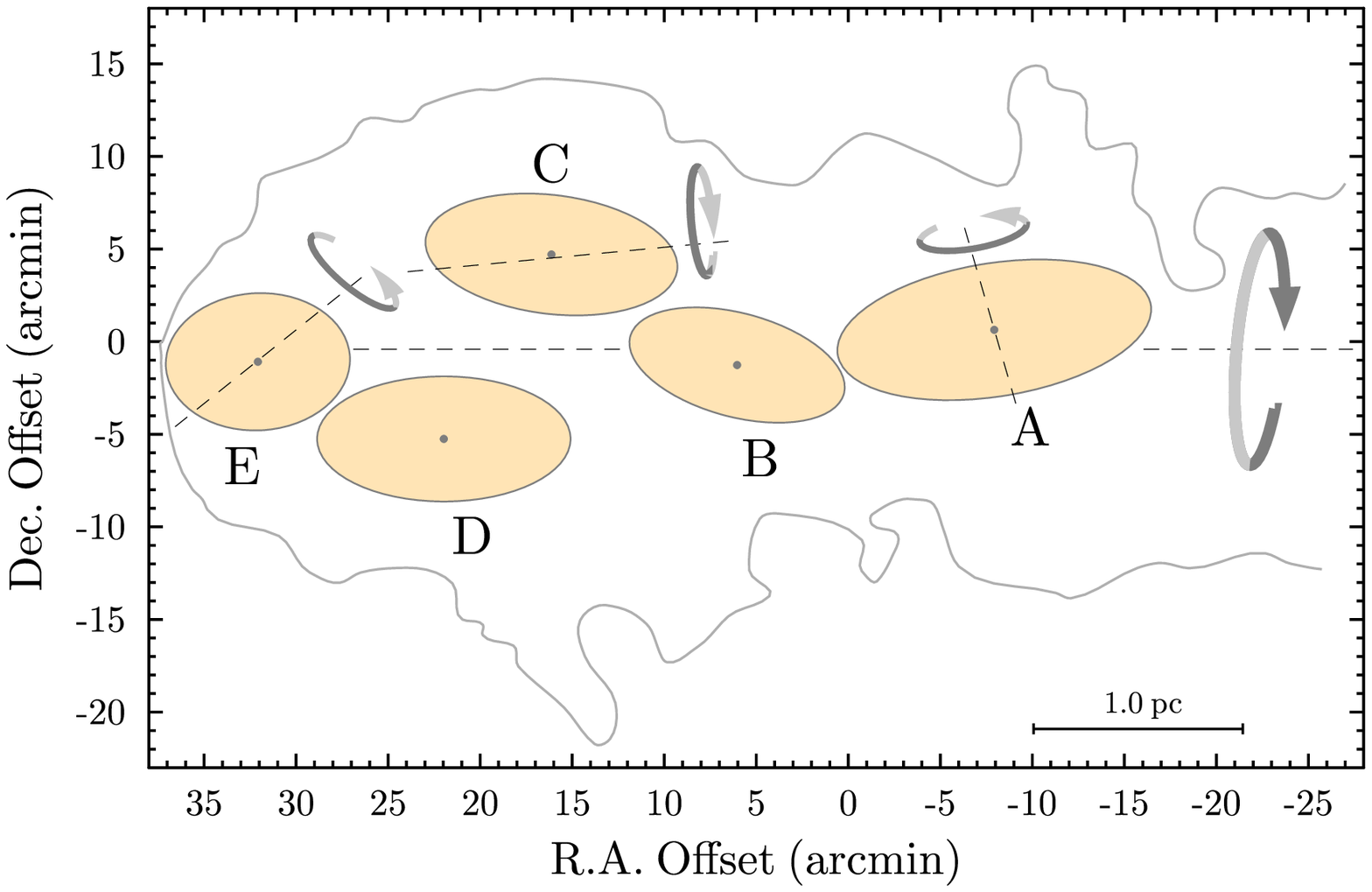,height=16.0cm,width=11.0cm}
\vspace{-6.0cm}
\caption[]{Morphology of a 
cometary-shaped cloud \object{L1251} as revealed in Sato \etal\ 1994.  
The contour shows the integrated $^{13}$CO(1-0) emission 
at the lowest level of 1.5 K~km~s$^{-1}$  (Fig.~2{\bf b} in Sato \etal\ 1994).
Five ellipses represent locations and angular sizes of five
C$^{18}$O dense cores (Tables 1 and 2 in S94).  
The rotation of the individual cores is shown by small-sized gray arc arrows. 
The global rotation of the $^{13}$CO cloud is indicated by the large-sized arc arrow.
For each arc arrow, denser gray color indicates the outer near side of the surface of the arc.
The dashed lines are the rotation axes. 
The (0,0) map position is R.A. = 22:31:02.3, Dec = 75:13:39 (J2000).
The adopted distance to the cloud is 300 pc.
}
\label{fg1}
\end{figure}

Figure~\ref{fg1} shows schematically cloud morphology with
ellipses representing locations and angular sizes of the five
C$^{18}$O dense cores. The boundary of the cloud (shown by a gray line) is set by
the integrated $^{13}$CO(1-0) emission at
the lowest level of 1.5 K~km~s$^{-1}$ (see Fig.~2{\bf b} in S94). 
The (0,0) map position is R.A. = 22:31:02.3, Dec = 75:13:39 (J2000),
which is fixed throughout the present paper.

The kinematic structure of the radial velocity field of \object{L1251} 
reveals a number of peculiar features. 
The $^{13}$CO(1-0) and C$^{18}$O(1-0) emission lines show
a highly ordered velocity gradient of ${\cal G} \approx -0.5$ \kmspc\, 
in the direction parallel to the minor axis of the cloud 
in the head of the cloud east of the offset R.A. $\sim 0'$ in Fig.~\ref{fg1}.
Since it is interpreted as a solid body rotation, this gradient corresponds to an
angular velocity of $\Omega_{\scriptscriptstyle {\rm CO}} \approx -2\times10^{-14}$ \rds.
The negative sign means that the radial velocity is more negative to the north; 
i.e., the velocity gradient is directed N$\to$S across the head.
This global motion is indicated in Fig.~\ref{fg1} 
by a large-sized arc arrow around the horizontal E--W axis (dashed line). 
In this respect \object{L1251} resembles rotating elephant trunks
found in star formation regions
(Pound \etal\ 2003; Hily-Blant \etal\ 2005; Gahm \etal\ 2006).
The cores A, B, and E disposed along this
axis have approximately the same radial velocity \VLSR $\approx -4.0$ \kms,
whereas for the northern core C and the southern core D  
\VLSR(C) $\approx -4.6$ \kms, and \VLSR(D) $\approx -3.8$ \kms.
(A typical uncertainty of \VLSR\ is $\sim 0.1$ \kms, S94.)

However, observations in \nhhh(1,1) revealed that
despite the N$\to$S global velocity gradient,
the northern core C exhibits a counter-rotation characterized by 
a highly ordered velocity gradient with
${\cal G}_{\rm C} \approx 1.7$ \kmspc, and position angle PA $\sim 30$\degr\ (E of N),
corresponding to the angular velocity of
$\Omega_{\scriptscriptstyle {\rm NH}_3} \approx 4\times10^{-14}$ \rds\
(see Fig.~6 in Levshakov \etal\ 2014). 
Taking into account that C-bearing molecules are usually distributed in the outer parts of the cores
(e.g., Tafalla \etal\ 2004), we conclude that the denser interior of this core traced by ammonia 
emission moves in a direction opposite to the global motion of the CO envelope.

For core A, Goodman \etal\, (1993)
reported an approximately east-west velocity gradient in \nhhh\
(${\cal G}_{\rm A} \approx 1.3$ \kmspc, and PA $\approx -77$\degr)
with the rotation axis oriented 
almost perpendicular to the horizontal (E--W) rotation axis of the dark cloud 
\object{L1251} (also T\'oth \& Walmsley 1996, hereafter TW96).

A more complex gas motion is observed in core E.
According to Goodman \etal\, (1993), 
who used spectral-line maps from Benson \& Myers (1989),
the velocity gradient direction in E is nearly orthogonal to the gradient direction in A 
(${\cal G}_{\rm E} \approx 3.5$ \kmspc, and PA $\approx -156$\degr). 
However, Caselli \etal\ (2002) mapped this core in \nnhp(1-0)
with an angular resolution of 54\arcsec\ 
(1.5 times the angular resolution of Benson \& Myers)
and found reverse velocity gradients in the eastern and western parts of the core 
suggestive of two counter-rotating adjacent clumps.
For these clumps taken together, the \nnhp\ tracer shows
${\cal G}_{\rm E} \approx 1.7$ \kmspc and PA $\approx -130$\degr\ (Caselli \etal\ 2002), meaning that
the directions of increasing velocity for both \nhhh\ and \nnhp\ are significantly correlated. 
Further single-dish and interferometric studies supported
very complex kinematics, including rapid rotation, infall, and outflow motions in the dense core E
(Lee \etal\ 2007). The directions of rotation and related rotation axes inferred from these studies 
are depicted in Fig.~\ref{fg1}. 
No information exists on kinematic processes in cores B and D, except for radial velocities
and linewidths of the $^{13}$CO(1-0) and C$^{18}$O(1-0) lines
measured in S94.

Here we continue our investigations of dense cores of the dark cloud \object{L1251}
in the \nhhh(1,1) and (2,2) inversion transitions with the Effelsberg 100-m
telescope\footnote{The 100-m telescope at Effelsberg/Germany is operated
by the Max-Planck-Institut f{\"u}r Radioastronomie on behalf of the
Max-Planck-Gesellschaft (MPG).}.  
In a previous paper (Levshakov \etal\ 2014), we mapped core C.
The present target is core A, which
is one of the ammonia emitters exhibiting some of the
narrowest (\Dv\ $\la 0.2$ \kms)\footnote{\Dv\ is the full width to half power (FWHP) value
throughout the paper.} 
lines ever observed (Jijina \etal\ 1999).
For this reason \object{L1251} was included in our list of targets 
used to validate the Einstein equivalence principle~-- local position invariance 
(Levshakov \etal\ 2010; Levshakov \etal\ 2013b).

\section{Observations}
\label{sect-2}

The ammonia observations were carried out with the Effelsberg 100-m telescope
in one session on April 22--27, 2014. 
The measurements were performed in the position-switching 
mode with the backend XFFTS (eXtended bandwidth FFTS) operating
at 100~MHz bandwidth and providing 32,768 channels
for each polarization. The resulting channel width was
$\Delta_{\rm ch} = 0.039$ \kms, but the true velocity resolution is 1.16 times
coarser (Klein \etal\ 2012).
The \nhhh\ lines were measured with a K-band high-electron mobility transistor (HEMT)
dual channel receiver, yielding spectra with a spatial resolution of $40''$ (FWHP) 
in two orthogonally oriented linear polarizations 
at the rest frequency of the $(J,K) = (1,1)$ and (2,2) lines, 
$f_{1,1} = 23694.495487$ MHz and $f_{2,2} = 23722.633644$ MHz (Kukolich 1967). 
Averaging the emission from both
channels, the typical system temperature (receiver noise and atmosphere)
is 100\,K on a main beam brightness temperature scale.

The mapping was done on a 40\arcsec\ grid.
The pointing was checked every hour by continuum cross scans of nearby continuum sources. 
The pointing accuracy was better than $5''$. The spectral line data were calibrated by means
of continuum sources with a known flux density. We mainly used G29.96--0.02 (Churchwell \etal\ 1990). 
With this calibration source, a main beam brightness temperature scale, \Tmb, can be established.
Since the main beam size ($40''$) is smaller than most core
radii ($>50''$) of our target, the ammonia emission couples
well to the main beam so that the \Tmb\ scale is appropriate.
Compensations for differences in elevation between the
calibrator and the source were $\la 20$\% and have not been taken
into account. Similar uncertainties of the main beam brightness
temperature were found from a comparison of spectra toward the same position taken on different dates.

The typical rms noise in our observations was 0.3~K per channel in main-beam brightness temperature units 
($150^{\rm s}$ ON and $150^{\rm s}$ OFF positions for one scan).
Some points were observed several times resulting in rms $\sim 0.15$~K.
For a characteristic line width of 0.2 \kms\ (see below), this yields typical
rms values of 0.05 K~km~s$^{-1}$, so that we choose
0.2 K~km~s$^{-1}$ as the lowest contour in the following.

\section{Results}
\label{sect-3}

The ammonia spectra were analyzed in the same way as in Levshakov \etal\ (2014).
There are no features with two velocity components.
The radial velocity, \VLSR,  the linewidth \Dv,
the optical depths \ta\ and \tb, the integrated ammonia
emission $\int$\Tmb$dv$, and the kinetic temperature \Tkin\  are all well-determined physical parameters, 
whereas the excitation temperature \Tex, the ammonia column density $N$(\nhhh), and the gas
density \nHH\ are less certain since they depend on the beam filling factor $\eta$,
which is not known for unresolved clumps. 
Assuming that $\eta$ ranges between $\eta_{\rm min}$ and 1, one can estimate limits
for \Tex, $N$(\nhhh), and \nHH, where a minimum value of \nHH\ 
is obtained by the choice of $\eta = 1$.
The corresponding equations are given in Appendix A in 
Levshakov \etal\ (2013a, hereafter L13).

The errors of the model parameters were estimated from the diagonal elements of the
covariance matrix calculated for the minimum of $\chi^2$. 
The error in \VLSR\ was also estimated independently by the $\Delta \chi^2$ method 
(Press \etal\ 1992).  When the two estimates differed, the larger error was adopted.
Table~\ref{tbl-1} illustrates that a typical uncertainty of \VLSR\ and \Dv\ 
at peak intensities of \nhhh(1,1) is $\sim 0.01$ \kms\ ($1\sigma$ confidence level, C.L.). 
Below we describe the obtained results in detail.

\subsection{Global gas morphology from \nhhh }
\label{sect-3-1}

\begin{figure*}[t]
\vspace{0.0cm}
\hspace{-0.5cm}\psfig{figure=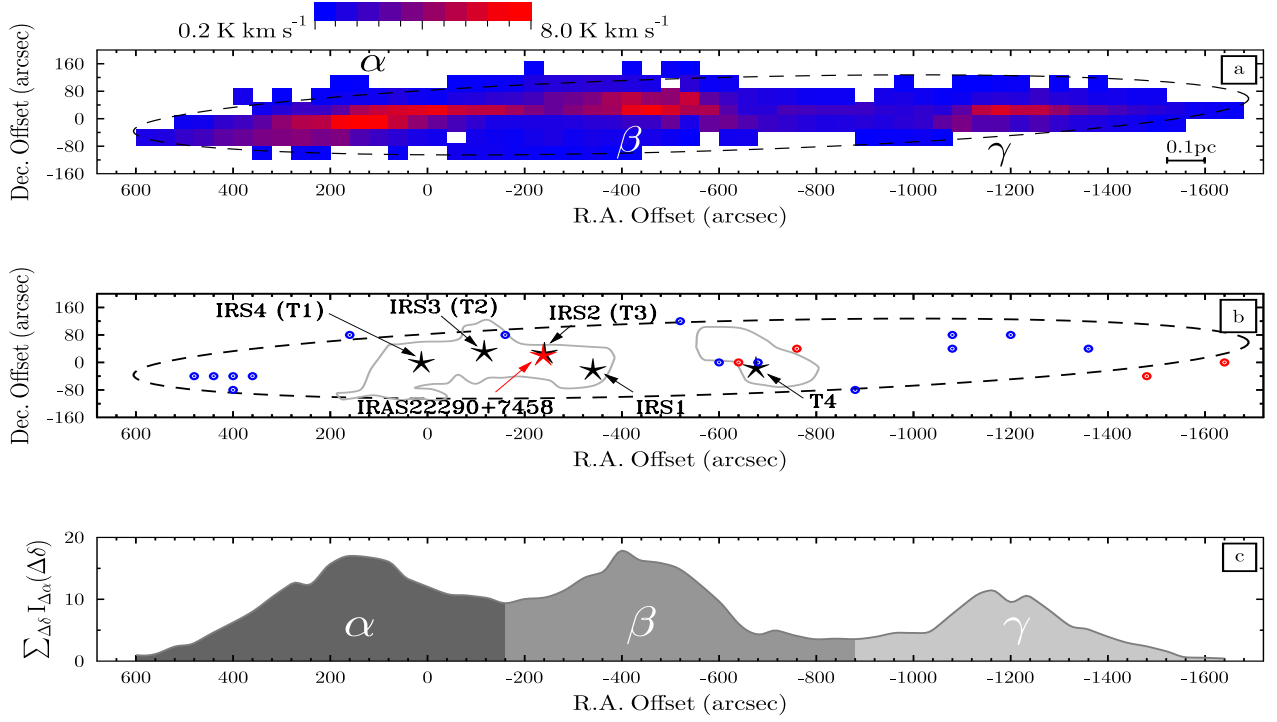,height=24.0cm,width=20.0cm}
\vspace{-10.0cm}
\caption[]{
({\bf a}) \nhhh(1,1) integrated intensity map ($\int$\Tmb$dv$ in units of K~\kms)
of the molecular core \object{L1251A}.
Each color box has a size of $40''\times40''$, and its 
bottom left corner corresponds to the telescope pointing.
The parameters of the \nhhh\ peaks $\alpha, \beta$, and $\gamma$
are given  in Table~\ref{tbl-1}.
The core exhibits a filamentary structure that can be enveloped by 
an ellipse (dashed line) with principal axes of 1140\arcsec\ and 89\arcsec.
({\bf b}) Schematic configuration of the filament and
peak positions of the ammonia cores T1--T4. 
The gray contours restrict ammonia emission to
the lowest level of the integrated \nhhh(1,1) emission
as in Fig.~3 in T\'oth \& Walmsley (1996).
The location of the IRS sources are from Lee \etal\ (2010).
The red star is the IR source detected by {\it IRAS}.
The filled blue and red circles outline areas of
the narrowest ammonia lines with FWHP line widths
of, respectively, $\Delta v \sim 0.2$ \kms, and $\sim 0.1$ \kms;
their offsets are listed in Table~\ref{tbl-1}.
({\bf c}) The sum over the integrated intensities for a given R.A.,
$\sum_{\Delta\delta} I_{\Delta\alpha}(\Delta\delta)$. 
The (0,0) map position and the distance to the source are as in Fig.~\ref{fg1}.
}
\label{fg2}
\end{figure*}

Our \nhhh\ observations cover the whole molecular core A and part of core B 
of \object{L1251} and consist of 300
measured positions, each of which is represented by a color box in the \nhhh(1,1) integrated intensity
map in Fig.~\ref{fg2}{\bf a}. The \nhhh\ map clearly reveals three peaks
that are labeled by $\alpha$, $\beta$, and $\gamma$.
Their parameters are given in Table~\ref{tbl-1}.
In Fig.~\ref{fg2}{\bf a}, the peak positions of cores B and A have offsets  
$(\Delta\alpha,\Delta\delta) = (362\arcsec, -76\arcsec)$ and
$(-474\arcsec, 38\arcsec)$, respectively.  Thus, the ammonia $\beta$ peak coincides with 
the C$^{18}$O peak A, whereas the C$^{18}$O peak B is slightly shifted from the ammonia $\alpha$ 
peak to the southeast.
For this reason we will call the whole structure traced by 
\nhhh\ emission as \object{L1251A}.

As mentioned above, the ammonia map was sampled at intervals of 40\arcsec,
which is half the grid size of 80\arcsec\ in TW96. 
The denser sampling resulted in a different apparent morphology of the \nhhh\ distribution.
For instance, TW96 identified four ammonia cores dubbed as ``T1'' to ``T4'' 
in decreasing R.A. direction. The first three T1--T3 have a common envelope, whereas T4
is separated (see Fig.~\ref{fg2}{\bf b}).
However, our results show that these two zones of ammonia emission
are not separated, but, in fact, all four cores belong to a central part of a continuous elongated structure 
that is much larger than the area of the \nhhh\ emission
mapped by TW96. The peak positions of cores T1-T4 are marked by black stars in Fig.~\ref{fg2}{\bf b}
for comparison with our observations. The known infrared sources are indicated as well.

The shape of the filament at the lowest level of 0.2 K~\kms\ 
can be approximated by an ellipse with
the angular sizes of the major (E-W) and minor (S-N) axes 
on the plane of the sky of about 38\arcmin\ and 3\arcmin.
The projected linear sizes of these 
axes are $\ell_1 \approx 3.3$ pc and $\ell_2 \approx 0.3$ pc; i.e.,
the aspect ratio is $\varepsilon \approx 11$. 

The \nhhh(1,1) integrated intensities $I(\Delta\alpha,\Delta\delta) = \int T_{\rm MB}dv$ form three condensations
around the $\alpha$, $\beta$, and $\gamma$ peaks, which are clearly seen in Fig.~\ref{fg2}{\bf c}
where we plot the sum over $\Delta\delta$ values, 
$\sum_{\Delta\delta} I_{\Delta\alpha}(\Delta\delta)$, 
at each fixed $\Delta\alpha$ as a function of R.A. 
The shape of this function, shown by gray in Fig.~\ref{fg2}{\bf c}, 
allows us to formally assign the boundaries for these condensations:
$-160\arcsec \la \Delta\alpha \leq 600\arcsec$ for $\alpha$,
$-880\arcsec \la \Delta\alpha \la -160\arcsec$ for $\beta$,
and $-1640\arcsec \leq \Delta\alpha \la -880\arcsec$ for $\gamma$.

\begin{figure*}[t]
\vspace{-2.0cm}
\hspace{-0.5cm}\psfig{figure=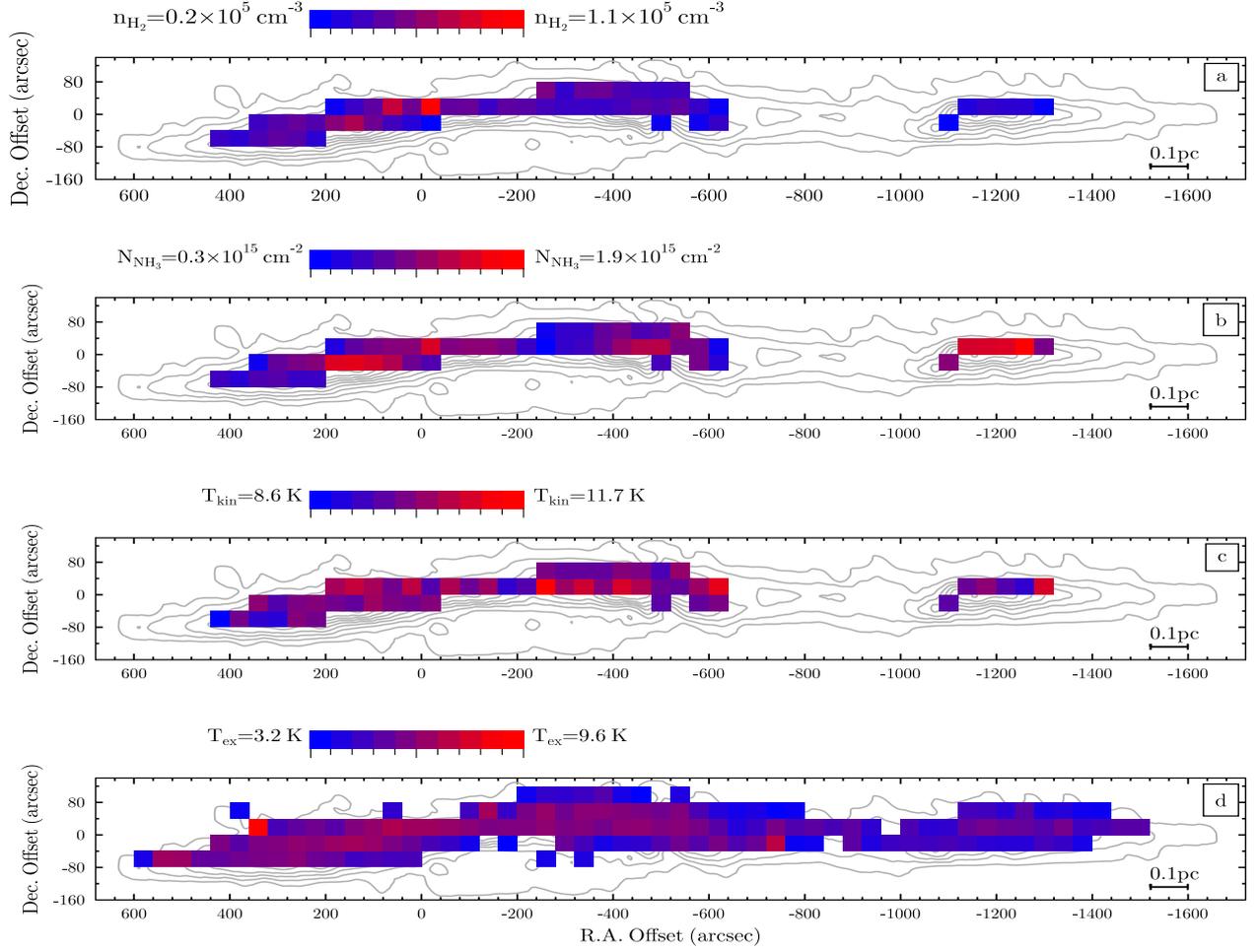,height=24.0cm,width=20.0cm}
\vspace{-7.0cm}
\caption[]{Distributions of the measured parameters in \object{L1251A}:
({\bf a})~--- the gas volume density, $n_{\scriptscriptstyle {\rm H_2}}$; 
({\bf b}) the \nhhh\ column density, $N_{\scriptscriptstyle {\rm NH_3}}$; 
({\bf c}) the kinetic temperature, \Tkin; 
and ({\bf d}) the excitation temperature, \Tex. 
Three parameters $n_{\scriptscriptstyle {\rm H_2}}$, $N_{\scriptscriptstyle {\rm NH_3}}$, and \Tkin\
were measured 
at the 54 positions where both the \nhhh(1,1) and (2,2) lines were observed, whereas \Tex\ was estimated
from the \nhhh(1,1) lines, which are not very weak.
Each color box has a size of $40''\times40''$ and its 
bottom left corner corresponds to the telescope pointing.
The contours are the \nhhh(1,1) integrated intensity map.
The starting point for the contour levels is 0.2 K~\kms; the increment is 1.0 K~\kms. 
The (0,0) map position is as in Fig.~\ref{fg1}.
We note that for the beam filling factor $\eta < 1$, certain parameters could be larger 
(see Sect.~\ref{sect-3} and \ref{sect-3-1} for details).
}
\label{fg3}
\end{figure*}

The thickness of the filament along the line of sight (the third axis $\ell_3$) can be estimated from
the measured ammonia column density, $N_{\scriptscriptstyle \rm NH_3}$, 
and the gas number density, $n_{\scriptscriptstyle \rm H_2}$, 
assuming a mean abundance ratio $[{\rm NH}_3]/[{\rm H}_2] = (4.6\pm0.3)\times10^{-8}$ 
found in dark clouds (Dunham \etal\ 2011). 
The measured values of $n_{\scriptscriptstyle \rm H_2}$ and $N_{\scriptscriptstyle \rm NH_3}$
over the ammonia map are shown in panels ({\bf a}) and ({\bf b}) in Fig.~\ref{fg3}.
Along the major axis $\ell_1$,  we observed both the \nhhh(1,1) and (2,2) transitions at 54 positions 
between the offsets $\Delta\alpha = 440\arcsec$ and $-1280\arcsec$, 
except for a gap $-1040\arcsec \leq \Delta\alpha \leq -640\arcsec$.  Averaging this dataset  we obtain
${\langle N\rangle}_{\scriptscriptstyle \rm NH_3} =
(1.03\pm0.05)\times10^{15}$ cm$^{-2}$ ($1\sigma$ C.L.)\footnote{Uncertainties
of the measured quantities correspond to $1\sigma$ C.L.} 
and  ${\langle n\rangle}_{\scriptscriptstyle \rm H_2} = (3.9\pm1.3)\times10^4$ cm$^{-3}$ 
under the assumption that the beam filling factor $\eta = 1$ 
(Eqs.~(A.19) and (A.21) in L13); i.e.,  
we observe a rather small ($\sim 5\%$) fluctuation in column densities along with
a significant variation ($\ga 30\%$) in gas number densities.
For instance, a twofold variation is found between $\alpha$ peak where 
$n^{\scriptscriptstyle \alpha}_{\scriptscriptstyle {\rm H_2}} = 7.1\times10^4$ \cmm\ 
and other peaks with
$n^{\scriptscriptstyle \beta}_{\scriptscriptstyle {\rm H_2}} = 3.8\times10^4$ \cmm\ and
$n^{\scriptscriptstyle \gamma}_{\scriptscriptstyle {\rm H_2}} = 3.3\times10^4$ \cmm.
Besides this, a sharp change in $n_{\scriptscriptstyle {\rm H_2}}$ is measured between 
the offsets (0\arcsec,0\arcsec) and $(0\arcsec,-40\arcsec)$ 
showing $n_{\scriptscriptstyle {\rm H_2}} = 6.8\times10^4$ \cmm\ and $1.7\times10^4$ \cmm, respectively.

The reason for such a high density fluctuation is not clear. 
It might be due to the used filling factor $\eta = 1$ and the presence of small clumps with linear sizes less
than 0.05 pc that are unresolved in our observations\footnote{A size of $\sim 0.1$ pc 
and an aspect ratio $\varepsilon \sim 2$ were found for a large
majority of low-mass cores mapped in \nhhh 
(e.g., Myers \etal\ 1991; Ryden 1996; Caselli \etal\ 2002; Kauffmann \etal\ 2008).}.
Shown in Fig.~\ref{fg3}{\bf d},
the excitation temperature \Tex, also estimated at $\eta = 1$ (Eq.~(A.10) in L13), 
may be inaccurate as well, since \Tex\ then corresponds to a minimum value.
With increasing \Tex\ (i.e., decreasing $\eta$) and, 
as a result, decreasing difference $(T_{\rm kin} -T_{\rm ex})$,
the estimate of $n_{\scriptscriptstyle {\rm H_2}}$ loses stability (Ho \& Townes 1983). 
In this difference, \Tkin\ is actually a fixed quantity since
the kinetic temperature is found to be almost constant over the whole filament (see Fig.~\ref{fg3}{\bf c}).
Its mean value is $\langle T_{\rm kin}\rangle = 10.12\pm0.08$ K.

For the mean abundance $[{\rm NH}_3]/[{\rm H}_2] = 4.6\times10^{-8}$, the ratio of 
${\langle N\rangle}_{\scriptscriptstyle \rm NH_3}$ and ${\langle n\rangle}_{\scriptscriptstyle \rm H_2}$ gives
the mean thickness of the filament of $\langle{\ell}_3\rangle \sim 0.2$ pc, which is comparable to $\ell_2$.
We note that for $\eta < 1$, $\langle{\ell}_3\rangle$ would become smaller.
Thus, if we use the sample means, then the filamentary structure traced by ammonia emission 
resembles a prolate ellipsoid of revolution (rod-like) with $\ell_1 \gg \ell_2 \ga \ell_3$.

\begin{figure*}[t]
\vspace{0.0cm}
\hspace{-0.5cm}\psfig{figure=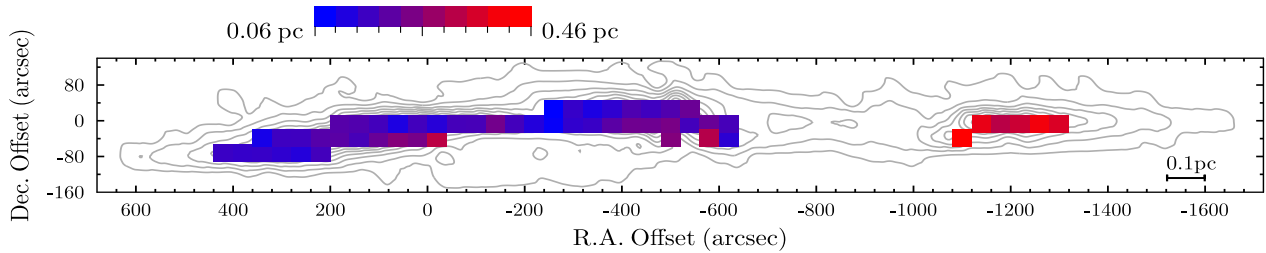,height=26.0cm,width=20.0cm}
\vspace{-17.0cm}
\caption[]{
Thickness along the line of sight through the core \object{L1251A}
estimated from the measured column densities $N_{\scriptscriptstyle {\rm NH_3}}$
and number densities $n_{\scriptscriptstyle {\rm H_2}}$ shown in Fig.~\ref{fg3}. 
The mean abundance ratio
of $[{\rm NH}_3]/[{\rm H}_2] = 4.6\times10^{-8}$ (Dunham \etal\ 2011)
is fixed for the whole filament.
The (0,0) map position is as in Fig.~\ref{fg1}.
We note that for the beam filling factor $\eta < 1$, the thickness is smaller 
(see Sects.~\ref{sect-3} and \ref{sect-3-1} for details).
}
\label{fg4}
\end{figure*}

However, the ammonia filament may have a more complex 
morphology if one compares individual gas densities at each point. 
Following this approach, we obtained local thickness
at different positions as displayed in Fig.~\ref{fg4}. 
It is seen that the linear dimension $\ell_3$ within the area occupied by 
the $\alpha$ and $\beta$ condensations is $\sim 0.2$ pc, whereas 
it is two times larger, $\sim 0.4$ pc, at the position of the $\gamma$ condensation.

It has to be noted that such calculations of $\ell_3$ are only legitimate
at gas densities below or near the critical density, 
$n_{\scriptscriptstyle {\rm H_2}} \leq n_{\rm cr}$, since otherwise
the observed line intensity is no longer unambiguously related to the gas density because
the collisionally induced transitions produce no photons. 
For \nhhh(1,1), the critical density at \Tkin\ = 10~K
is $n_{\rm cr} = 3.90\times10^3$ \cmm\ (Maret \etal\ 2009), 
which is an order of magnitude lower than the measured mean value of
${\langle n}\rangle_{\scriptscriptstyle {\rm H_2}}$. 
Therefore, the estimates of $\ell_3$ must be taken with some caution.
Nevertheless, a sheet-like geometry of \object{L1251A} with $\ell_3 \sim \ell_1$ 
may be excluded since it would produce too low a gas number density,
$n_{\scriptscriptstyle {\rm H_2}} \sim 2\times10^3$ \cmm~--- 
an order of magnitude lower than the measured values of \nHH. 
An $\eta < 1$ would make $\ell_3$ even shorter.

In what follows, we assume that the filament has a rod-like geometry
consisting of two cylinders with different cross-sections: 
a circular one with diameter 0.2 pc for $\alpha$ and $\beta$,
and an elliptical cylinder with the major and minor axes of 0.4 pc and 0.2 pc for $\gamma$. 
Their corresponding linear sizes are 2.2 pc and 1.1 pc.
The mean gas number densities are
${\langle n \rangle}_{\alpha\beta} = (4\pm2)\times10^4$ \cmm\ (sample size = 48), and
${\langle n \rangle}_{\gamma} = (2.6\pm0.4)\times10^4$ \cmm\ (sample size = 6).

It is interesting to compare our results with previous ones.
According to TW96, the angular size (E-W) of the combined cores T1--T3 is $\sim7\arcmin$,
which is in line with the ammonia map of Goodman \etal\ (1993). Together with T4, the total
size of the T1--T4 complex in the E-W direction is about 15\arcmin, which is
more than two times shorter as compared with $\ell_1$.
In the orthogonal direction (S-N), we observe the same extension of the ammonia map
as reported by Goodman \etal\ and TW96.

\begin{figure}[t]
\vspace{0.0cm}
\hspace{-0.5cm}\psfig{figure=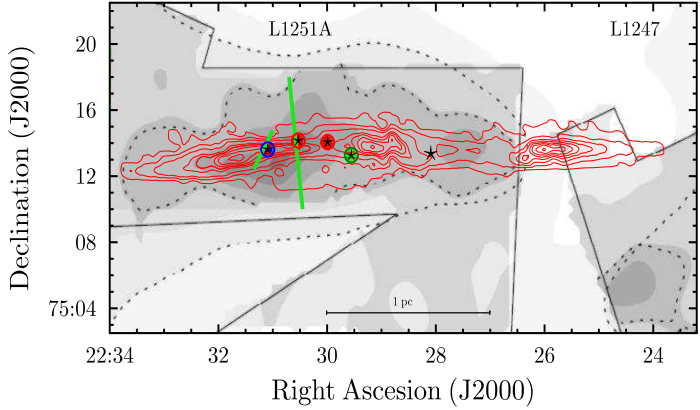,height=11.0cm,width=10.0cm}
\vspace{-1.0cm}
\caption[]{
Portion of Fig.~17 from Kirk \etal\ (2009) with extended structure toward
\object{L1251A} and \object{L1247}.
The gray scale shows the distribution of visual extinction 
as derived from the Digitized Sky Survey (Dobashi \etal\ 2005), and
the superimposed higher resolution extinction maps (outlined by solid straight black lines)
observed with the {\it Spitzer} Infrared Array Camera (IRAC; 3.6-8.0 $\mu$m) 
and Multiband Imaging Photometer (MIPS; 24-160 $\mu$m). 
Two dashed lines refer to Dobashi $A_V = 1$ mag and {\it Spitzer} $A_V = 5$ mag. 
The colored markers show the location and spectral type of the 
young stellar object (YSO) candidates: red/green/blue for Class I/Flat/Class II. 
The black stars are the same markers as in Fig.~\ref{fg2}{\bf b}.
The \nhhh(1,1) integrated intensity map, as in Fig.~\ref{fg4}, is denoted by
the red contours.
The approximate orientation and linear size of 
a jet extending about 10\arcmin\ and a small bipolar nebula 
to the east of the jet (solid green lines) are copied from
Fig.~1 in Lee \etal\ (2010).
}
\label{fg5}
\end{figure}

The previous ammonia observations did not reveal the $\gamma$ condensation 
which is partly overlapping the nearby dark cloud \object{L1247} (Fig.~\ref{fg5}). 
The maximum of the \nhhh(1,1) integrated intensity  
lies in the region with minimum visual extinction just between \object{L1251A}
and \object{L1247}.
Thus, approximately three fourths of
the ammonia map coincides with the most prominent filament evident in the dust emission
(dark gray area with $A_V \ga 5$ mag in Fig.~\ref{fg5}), whereas one fourth follows 
a weaker dust emission with $A_V \sim 1$ mag.

The question then arises as to how it is possible that in the $\gamma$ region the
maximum of the ammonia emission originates in areas with minimal visual extinction.
This is perhaps due to small scale clumping with the \nhhh\ gas 
providing high visual extinction but not covering the entire area,
hinting at an $\eta < 1$ for this region. 
Possibly, $\eta$ is smaller in the $\gamma$ fragment than in $\alpha \beta$.
In Sect.~\ref{sect-3-3}, we discuss that the average H$_2$ density is well below what is deduced from
\nhhh. This is also consistent with small scale clumping.

The comparison with the C$^{18}$O(1-0) map, 
which has the major (E$\to$W) and minor (S$\to$N) axes 
of 17\arcmin~$\times$~7\arcmin\ (Fig.~2 in S94), 
shows that in general both \nhhh\ and C$^{18}$O trace each other,
but the former is more concentrated on the horizontal E--W axis. 
This is not surprising since the C$^{18}$O(1-0)
transition becomes excited at densities an order of magnitude lower than needed for \nhhh(1,1).
However, an unexpected result is that a rather large area of ammonia emission from the western 
part of \object{L1251A} ($\gamma$ condensation in Fig.~\ref{fg2}{\bf a}) 
is not seen in C$^{18}$O: this part of \object{L1251} was mapped only in $^{13}$CO(1-0) (Fig.~\ref{fg1}) 
and in dust emission (Fig.~\ref{fg5}).  The $^{13}$CO ``tail'' of \object{L1251} 
west of the offset R.A. $\sim 0'$ in Fig.~\ref{fg1} has an angular size
of $\sim 27\arcmin \times 20\arcmin$, which is much larger than 
regions with visual extinction $A_V \sim 5$ mag and the similarly extended C$^{18}$O emission.

Considering the spatial distributions of these emitters, we see that the tail 
of \object{L1251A}
has a cocoon-like morphology with radial stratification from the
outer to the innermost layers seen in $^{13}$CO, C$^{18}$O, dust emission, and \nhhh. 
The radius of the first region 
projected on the plane of the sky  
is $R_1 \sim 0.9$ pc ($^{13}$CO), the second $R_2 \sim 0.3$ pc (C$^{18}$O and dust),
and the third $R_3 \sim 0.13$ pc (\nhhh).
The ambient medium traced by $^{12}$CO emission (the outermost envelope) 
has a very large angular size and irregular morphology as displayed
in Fig.~1 in Grenier \etal\ (1989).

In the next section, we show that all three ammonia
condensations $\alpha$, $\beta$, and $\gamma$ 
are involved in a common global motion of the filamentary structure
with some intrinsic peculiarities.

\subsection{Global gas kinematics from \nhhh }
\label{sect-3-2}

\begin{figure*}[t]
\vspace{0.0cm}
\hspace{-0.5cm}\psfig{figure=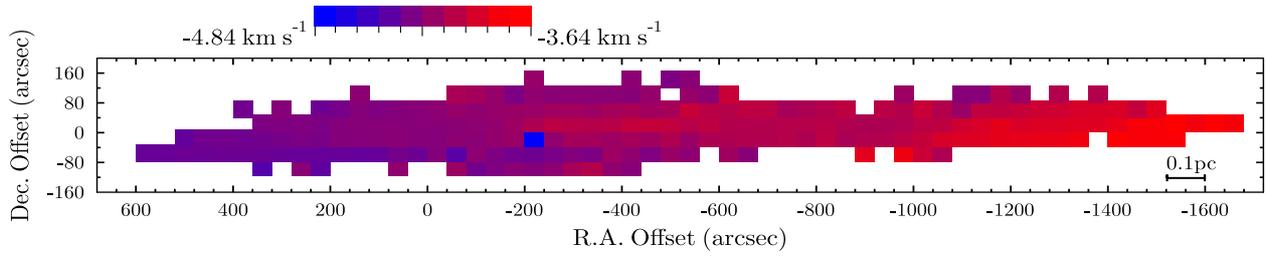,height=26.0cm,width=20.0cm}
\vspace{-17.0cm}
\caption[]{
Same as Fig.~\ref{fg4} but for the \nhhh(1,1) radial velocity field (\VLSR).
}
\label{fg6}
\end{figure*}

The \nhhh\ radial velocity map is shown in Fig.~\ref{fg6}.
In our analysis of ammonia spectra, all hyperfine structure (hfs) components of the \nhhh(1,1) and (2,2) transitions
were fitted simultaneously to determine the total optical
depth, $\tau_{\rm tot}$, in the respective inversion transition, the LSR velocity
of the line, \VLSR, the intrinsic full-width at half power linewidth,
\Dv\ (FWHP), for individual hf components, and the amplitude, ${\cal A}$
(see Appendix A in L13).  For unsaturated lines ($\tau_{\rm hf} \la 1$),
the velocity \VLSR\ represents an intensity-weighted average centroid velocity
along the line of sight through the cloud.

If the cloud rotates as a solid body, \VLSR\ is independent of 
distance along the line of sight, and linearly dependent on the coordinates
in the plane of the sky (e.g., Goodman \etal\ 1993; Belloche 2013).
Then, the cloud rotation velocity field is defined by the equation
\begin{equation}
{\vec V} = \vec{\Omega} \times {\vec R}\ ,
\label{Eq1}
\end{equation}
where $\Omega$ is the angular velocity, and $R$ the distance to the rotation axis.
This means that a cloud in solid-body rotation should demonstrate
a linear gradient, $\nabla$\VLSR, across the surface of a map,
orthogonal to the rotation axis $\vec{\Omega}$. 
This is just the case of \object{L1251A} where
the velocity gradient in the direction E$\to$W along the major axis
can be easily identified by eye (Fig.~\ref{fg6}).

\subsubsection{The whole filament, 
$-1640\arcsec \leq \Delta\alpha \leq 600\arcsec$
}
\label{sect-3-2-1}

The \VLSR\ values are systematically increasing with decreasing R.A. (E$\rightarrow$W)
on the total linear scale of the filament $\ell_1 \approx 3.3$ pc.
As it follows from Fig.~\ref{fg6}, the velocity difference between the
eastern and western edges of the filament is $\Delta V \approx 0.7$ \kms,
meaning that the gradient becomes $\nabla$\VLSR$ \approx 0.2$ km~s$^{-1}$~pc$^{-1}$.

\begin{figure}[t]
\vspace{0.0cm}
\hspace{-2.0cm}\psfig{figure=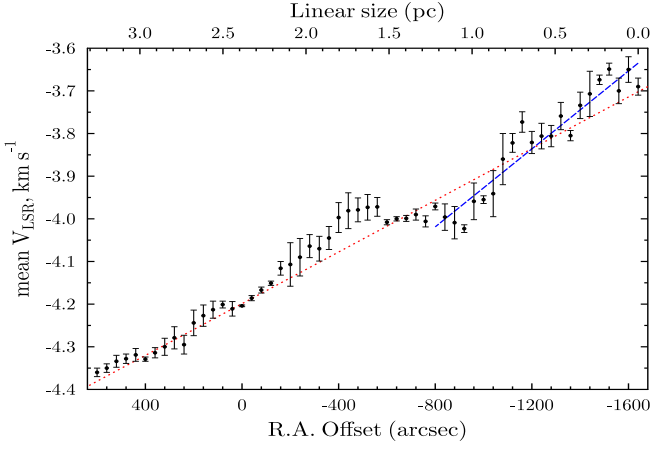,height=16.0cm,width=13.0cm}
\vspace{-7.0cm}
\caption[]{
Position-velocity diagram of \nhhh(1,1) along the direction
of the velocity gradient (E $\to$ W).
At each offset $\Delta\alpha$,
the measured centroid velocities (\VLSR) are averaged 
across the filament in the direction S$\to$N.
The error bars (the standard deviation of the mean)
result from this averaging.
The different positions were weighted inversely proportional to the variance.
The linear regression for the total range is shown by red:
$V_0 = -4.199(1) - 0.000303(3)\Delta\alpha$ (\kms) (here $\Delta\alpha$ is in arcsec).
For the $\alpha\beta$ condensation $(-480\arcsec \leq \Delta\alpha \leq 600\arcsec)$,
the regression line is practically the same.
For the $\gamma$ condensation $(-1640\arcsec \leq \Delta\alpha \leq -800\arcsec)$,
it is shown by a blue line with
$V_0 = -4.38(1) - 0.00046(1)\Delta\alpha$ (\kms).
The number in parentheses is the one-sigma ($1\sigma$) uncertainty in the last digit
of the given value.
The (0,0) map position is as in Fig.~\ref{fg1}.
}
\label{fg7}
\end{figure}

The same gradient can be found from the regression analysis shown in Fig.~\ref{fg7},
where a functional dependence 
of spatial averages of $\langle$\VLSR$\rangle$ on R.A. is depicted.
At each offset $\Delta \alpha$, the weighted mean value of $\langle$\VLSR$\rangle$ was found from
averaging along the S$\to$N direction with weights inversely proportional to the
uncertainties of the measured line-of-sight velocities.
The regression line (shown by red) is defined by 
\begin{equation}
V_0 = -4.199(1) - 0.000303(3)\Delta\alpha\ ({\rm km~s}^{-1})\ ,
\label{Eq2}
\end{equation}
which corresponds to $\nabla$\VLSR$ = 0.208\pm0.003$ km~s$^{-1}$~pc$^{-1}$. 
Here, the numbers in parentheses are $1\sigma$ errors on the last digit, 
and $\Delta\alpha$ is counted in arcsec.

It follows from Eqs.~(\ref{Eq1}) and (\ref{Eq2}) that 
the angular velocity of the whole filament around the vertical axis (S$\to$N)
is $\Omega^{\scriptscriptstyle\rm SN}_{\rm tot} \approx 7.0\times10^{-15}$ \rds,
and the period for a complete rotation 
$t^{\scriptscriptstyle\rm SN}_{\rm tot} \approx 3\times10^{7}$ yr.

\subsubsection{The $\alpha$ and $\beta$ kinematic fragments,
$-480\arcsec \leq \Delta\alpha \leq 600\arcsec$
}
\label{sect-3-2-2}

Figure~\ref{fg7} demonstrates that radial velocities increase gradually with R.A.
over the area occupied by the $\alpha$ and $\beta$ condensations from the offset
$\Delta\alpha = 600\arcsec$ to $\Delta\alpha \sim -480\arcsec$.
The regression line for this kinematic fragment
is practically the same as for the whole filament:
\begin{equation}
V_0 = -4.197(2) - 0.000305(7)\Delta\alpha\ ({\rm km~s}^{-1})\ . 
\label{Eq3}
\end{equation}
Its angular velocity around the vertical axis (S$\to$N) is
$\Omega^{\scriptscriptstyle\rm SN}_{\alpha\beta} \simeq \Omega^{\scriptscriptstyle\rm SN}_{\rm tot}$.

It is noteworthy that
the gas in this area exhibits an additional spinning around the horizontal axis (E$\to$W), 
which is revealed in a systematic difference between the radial velocities
toward the northern and southern ridges of the filament. 
To specify this effect numerically, we considered the probability density function (pdf)  
of line centroid velocity fluctuations 
$\Delta V = V_{\scriptscriptstyle \rm LSR} - V_0$ 
calculated at each offset position on the map of the $\alpha\beta$ fragment.
Here, $V_0$ is a local systemic velocity at a given $\Delta\alpha$, 
which is defined by (\ref{Eq3}).
The obtained pdf is shown by the blue histogram (approximated by a Gaussian) in Fig~\ref{fg8}{\bf a}. 
It consists of three peaks: the central component with $\Delta V = 0$ \kms, and the two lateral components 
with $\Delta V_{\scriptscriptstyle \rm N,S} \approx \pm 0.11$ \kms.
In this case, the positive northern component $\Delta V_{\scriptscriptstyle \rm N}$
indicates that the line-of-sight velocity is higher to the north, 
whereas the negative southern component $\Delta V_{\scriptscriptstyle \rm S}$ 
shows that \VLSR\ is predominantly directed to the observer.

The velocity difference between these two filament ridges is $\sim 0.22$ \kms, which gives 
the gradient $\nabla$\VLSR$ \sim 0.7$ km~s$^{-1}$~pc$^{-1}$ in the direction S$\to$N.
The corresponding angular velocity $\Omega^{\scriptscriptstyle\rm EW}_{\alpha\beta}$
has the opposite sign to the rotation of the CO envelope   
($\Omega_{\scriptscriptstyle {\rm CO}} \approx -2\times10^{-14}$ \rds)
and equals
$\Omega^{\scriptscriptstyle\rm EW}_{\alpha\beta} \approx 2\times10^{-14}$ \rds\
($t^{\scriptscriptstyle\rm EW}_{\alpha\beta} \approx 1\times10^{7}$ yr).

\subsubsection{The bar,
$-800\arcsec \leq \Delta\alpha \leq -480\arcsec$
}
\label{sect-3-2-3}

Figure~\ref{fg7} shows that near the geometric center of the filament 
in the range $-800\arcsec \leq \Delta\alpha \leq -480\arcsec$, 
the mean radial velocity between the adjoining points 
is held constant: 
$\langle V_{\scriptscriptstyle \rm LSR}\rangle = -3.997\pm0.004$ \kms. 
This kinematic fragment is also characterized by an extremely low linewidth of ammonia lines,
\Dv\ $\sim 0.1 - 0.2$ \kms\ (blue and red filled circles in Fig.~\ref{fg2}{\bf b}),
and decreasing gas density 
$n_{\scriptscriptstyle {\rm H_2}}$ from $4\times10^4$ \cmm\ to $2\times10^4$ \cmm\
with changing $\Delta\alpha$ from $-480\arcsec$ to $-600\arcsec$ along the cut 
$\Delta\delta = 0\arcsec$.
Beyond this interval, between $\Delta\alpha = -640\arcsec$ and $-1040\arcsec$, we did not
detect the second ammonia transition \nhhh(2,2), and, therefore, 
the gas number density was not estimated. 
The same tendency is observed for the ammonia column density:
$N$(\nhhh) decreases from $1.5\times10^{15}$ \cm\ to $0.5\times10^{15}$ \cm\ 
within the interval $-600\arcsec \leq \Delta\alpha \leq -480\arcsec$.

In what follows, we refer to this fragment of the filament as a ``bar''.
Figure~\ref{fg2}{\bf a} shows that it starts at the western edge of the $\beta$ condensation and
its projected linear size is $\ell_{\rm bar} \approx 0.5$ pc.

\subsubsection{The $\gamma$ kinematic fragment, 
$-1640\arcsec \leq \Delta\alpha \leq -800\arcsec$
}
\label{sect-3-2-4}

The regression line of the $\gamma$ kinematic fragment, 
ranging from $\Delta\alpha = -800\arcsec$ to $-1640\arcsec$ 
(shown by blue in Fig.~\ref{fg7}), is defined by 
\begin{equation}
V_0 = -4.38(1) - 0.00046(1)\Delta\alpha\ ({\rm km~s}^{-1})\ , 
\label{Eq4}
\end{equation}
i.e., $\nabla$\VLSR$ = 0.316\pm0.007$ km~s$^{-1}$~pc$^{-1}$. 
Thus, the western part of the filament rotates around the vertical axis (S$\to$N)
with a slightly higher angular velocity
$\Omega^{\scriptscriptstyle\rm SN}_\gamma \approx 1.0\times10^{-14}$ \rds\ than
$\Omega^{\scriptscriptstyle\rm SN}_{\rm tot}$.

\begin{figure}[t]
\vspace{0.0cm}
\hspace{0.0cm}\psfig{figure=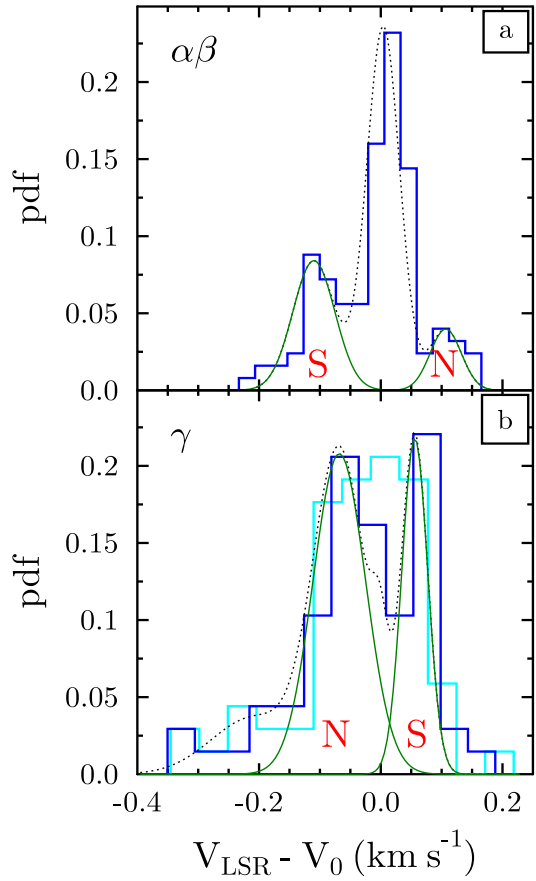,height=18.0cm,width=13.0cm}
\vspace{-4.0cm}
\caption[]{
Histograms are the probability density functions (pdfs) of line centroid velocity fluctuations
$\Delta V = V_{\scriptscriptstyle \rm LSR} - V_0$, and the smooth curves are their approximations
by Gaussians. Here $V_0$ is the systemic velocity component
defined by the regression lines shown in Fig.~\ref{fg7}.
Panels ({\bf a}) and ({\bf b}) represent the $\alpha\beta$ (red regression line in Fig.~\ref{fg7})
and $\gamma$ (blue regression line in Fig.~\ref{fg7})
condensations. The lateral peaks labeled by the letters 
``N''  and ``S'' correspond to the northern and southern ridges
of the filament. Note the opposite chirality of the angular
rotation around the horizontal axis E$\to$W 
(see Sects.~\ref{sect-3-2-2} and \ref{sect-3-2-4} for details).
}
\label{fg8}
\end{figure}

Toward the $\gamma$ condensation, we also reveal an additional systemic motion around 
the axis E$\to$W:  
the radial velocity is more negative to the north than what is observed from
the global rotation of the CO envelope. 
Applying the same procedure as for the $\alpha\beta$ fragment,
we calculated pdfs of the residuals $\Delta V$
using the two definitions of the systemic velocity $V_0$ given by 
Eqs.~(\ref{Eq2}) and (\ref{Eq4}). 
The resulting distribution functions are shown by the cyan and blue
histograms in Fig.~\ref{fg8}{\bf b}, respectively.
It is seen that the definition (\ref{Eq2}) based on the total dataset provides
unresolved lateral peaks and a broad central component,
whereas the regression line (\ref{Eq4}) allows us to partly resolve the lateral peaks.

It is remarkable that in this case we 
observe spinning with opposite chirality (dextral vs sinistral): 
the northern peak is shifted to negative 
line-of-sight velocities, and the southern~--- to positive velocities. 
Because of an asymmetric shape of the northern
peak, it was approximated by a convolution of
two Gaussians shown by the dotted curve in Fig.~\ref{fg8}{\bf b}. 
Their barycenter is
$\Delta V_{\scriptscriptstyle \rm N} \approx -0.089$ \kms, whereas
the center of the southern component is $\Delta V_{\scriptscriptstyle \rm S} \approx 0.056$ \kms.
This leads to a velocity difference between the northern and southern ridges of the filament
of $\sim -0.15$ \kms and
a gradient $\nabla$\VLSR$ \sim -0.5$ km~s$^{-1}$~pc$^{-1}$.
The corresponding angular velocity of the $\gamma$ fragment
around the axis E$\to$W is 
$\Omega^{\scriptscriptstyle\rm EW}_\gamma \approx -2\times10^{-14}$ \rds\
($t^{\scriptscriptstyle\rm EW}_\gamma \approx 1\times10^{7}$ yr).

Thus, here we find two kinematic fragments of opposite chirality 
when the filament is viewed from the eastern footpoint
along the axis E$\to$W:
the fragment $\gamma$ spins clockwise, whereas the fragment $\alpha\beta$ rotates counterclockwise. 
The filament chirality directly indicates the magnetic field helicity 
of two types, negative and positive (e.g., Martin 1998).

It should be emphasized
that all measured periods exceed the age of the supernova remnant considerably, 
$t_{\scriptscriptstyle\rm SNR} \sim 4\times10^4$ yr, mentioned in Sect.~\ref{sect-1}.
Since dark clouds evolve slowly, this may imply that the dynamical stage of \object{L1251A} 
was triggered by another supernova exploded in the past $10^{6-7}$ yr in
this region, as suggested in S94.  A similar lifetime of $\sim 10^6$ yr 
for starless cores with average volume density $n \sim 10^4$ \cmm\ 
was estimated by Lee \& Myers (1999), among others.

\subsubsection{Smoothness of the velocity field }
\label{sect-3-2-5}

The distribution of the centroid velocities over the map of 
the filament (Fig.~\ref{fg6}) reveals only one offset position
$(\Delta\alpha,\Delta\delta) = (-200\arcsec,-40\arcsec)$ (blue square)
with a peculiar velocity \VLSR\ = $-4.84(3)$ \kms\ which
deviates noticeably from velocities measured at the 
neighboring positions. For instance, 
\VLSR$(-200\arcsec,-80\arcsec) = -4.32(3)$ \kms,
\VLSR$(-200\arcsec,0\arcsec) = -4.019(7)$ \kms,
or
\VLSR$(-240\arcsec,0\arcsec) = -3.961(7)$ \kms,
which gives a break 
of $|\Delta V| \sim 0.5-0.9$ \kms\ on a scale $\sim 0.05$ pc. 
The nature of the outlier at ($-200$\arcsec,$-40$\arcsec) is not clear
and requires further exploration.
We note that this peculiar velocity is not a product
of measurement errors but is connected with other peculiarities
outlined further below in the same section.

\begin{figure}[t]
\vspace{0.0cm}
\hspace{0.0cm}\psfig{figure=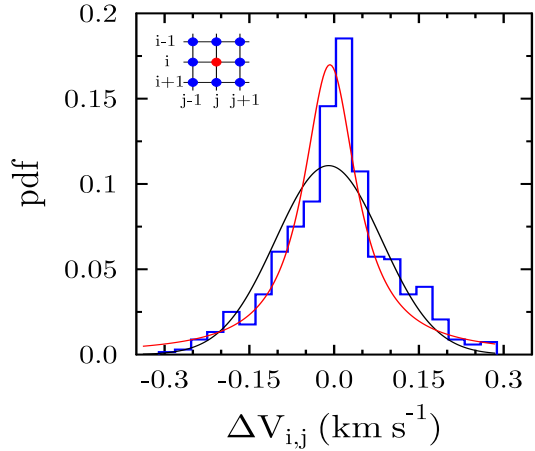,height=16.0cm,width=13.0cm}
\vspace{-8.0cm}
\caption[]{Histogram is 
the probability density function (pdf) of line centroid velocity fluctuations 
between neighboring positions
over the whole \nhhh\ map of the filament \object{L1251A} except for the  
outlier at the offset $(\Delta\alpha,\Delta\delta) = (-200$\arcsec,$-40$\arcsec).
In the upper left hand corner, the grid with color points 
illustrates the calculation of $\Delta V_{i,j}$:
for a current position $(i,j)$ (marked by red), the velocity differences are 
calculated between each blue point and the central red point.
The total number of different pairs is 680.
The histogram is approximated by a Gauss (black) and a Lorentz (red) function
with the widths (FWHP) of $\Delta v_{\scriptscriptstyle \rm G} = 0.22$ \kms\ and
$\Delta v_{\scriptscriptstyle \rm L} = 0.11$ \kms.
}
\label{fg9}
\end{figure}

At all other offsets, \VLSR\ changes smoothly from point to point. 
The variance of centroid velocity fluctuations at a given spatial lag
can be estimated from velocity differences between neighboring positions.
The histogram in Fig.~\ref{fg9} shows the pdf of such differences
at the lags $40\arcsec$ and $\sqrt{2}\cdot40\arcsec$ for the whole filament \object{L1251A}
except for the outlier at the offset ($-200$\arcsec,$-40$\arcsec).
For a given position $i,j$ (shown by red on the grid in the top lefthand corner of Fig.~\ref{fg9}),
the velocity differences $\Delta V_{i,j}$ are calculated between each blue point on the grid
and the central red point.
The total sample size consists of $p = 680$ pairs of these neighboring points.
The velocity difference pdf was approximated by Gauss (black curve) and Lorentz (red curve)
functions with the widths (FWHP) of $\Delta v_{\scriptscriptstyle \rm G} \approx 0.22$ \kms\ and
$\Delta v_{\scriptscriptstyle \rm L} \approx 0.11$ \kms.
The rms turbulent velocity at a lag $\sim 0.05-0.08$ pc  is equal to 
$\sigma_{\scriptscriptstyle \rm G} \approx 0.096$ \kms.
The value of $\sigma_{\scriptscriptstyle \rm G}$ depends weakly on the lag size.
For instance, two times and three times larger lags
provide $\sigma_{\scriptscriptstyle \rm G} \approx 0.129$ \kms\ $(p = 524)$ and
$\sigma_{\scriptscriptstyle \rm G} \approx 0.131$ \kms\ $(p = 381)$, respectively.
A similar weak dependence of the non-thermal velocity 
$\Delta v_{\scriptscriptstyle \rm NT}$ on radius,
$\Delta v_{\scriptscriptstyle \rm NT} \propto R^{0.09\pm0.05}$,
was noted by Goodman \etal\ (1998) for ammonia gas in \object{L1251A}. 

It is noteworthy that the Lorentz distribution in Fig.~\ref{fg9}
is in better concordance with the histogram, and its width
is narrower than that of the Gaussian. 
This may mean that the central part of the filament is organized into a   
coherent structure akin those found in some star-forming dense cores (e.g., Goodman \etal\ 1998;
Caselli \etal\ 2002; Pineda \etal\ 2010).
The size scale at which turbulent motions become coherent, 
the  ``coherence length'', is usually deduced from
the ``cutoff'' wavelength
below which Alfv\'en waves cannot propagate because of the
neutral medium opacity depending on the magnetic field strength,
$B$, the ionization fraction, $x_{\rm e}$, and the gas volume density, $n$.
For the typical values observed in dark clouds ($B \sim 10\mu$G, $x_{\rm e} \sim 10^{-8}$,
$n \sim 10^5$ \cmm), the coherence length is about 0.01 pc 
(e.g., Caselli \etal\ 2002).
Given the values of $\ell_1, \ell_2$, and $\ell_3$, there must be a large number
of coherent cells along each axis of the filament \object{L1251A,} which
may result in a non-Gaussian shape of 
the general velocity pdf (Fig.~\ref{fg9}). 

\begin{figure*}[t]
\vspace{0.0cm}
\hspace{-0.5cm}\psfig{figure=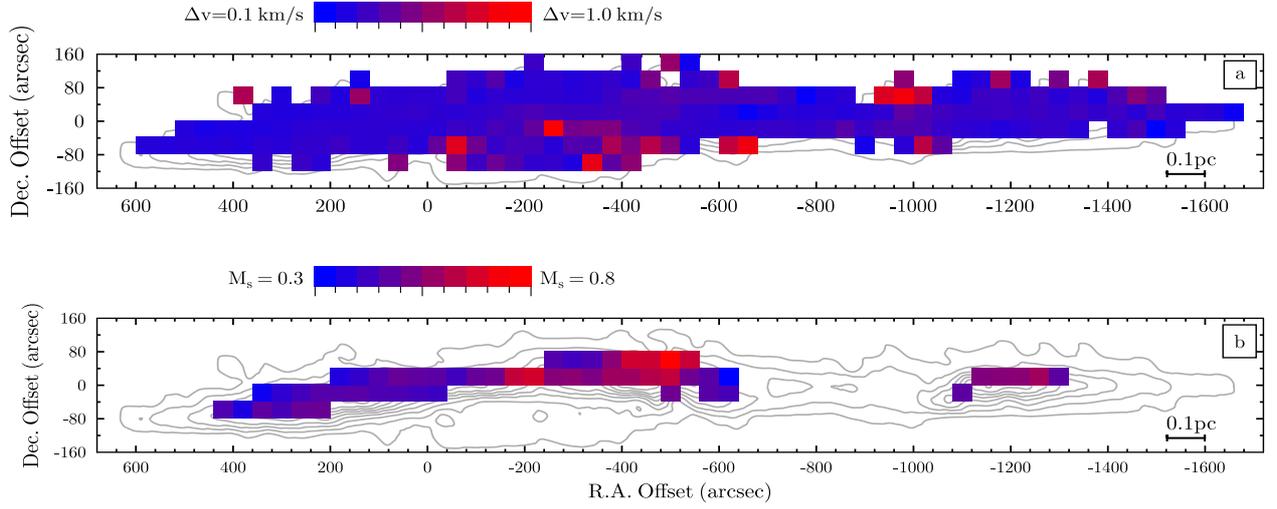,height=26.0cm,width=20.0cm}
\vspace{-14.0cm}
\caption[]{
({\bf a}) Linewidths, $\Delta v$, of the ammonia (1,1) lines (FWHP) measured toward 
the core \object{L1251A}.
({\bf b}) Mach numbers, $M_s$, of non-thermal motions
at the 54 positions where \Tkin\ was measured directly from 
the relative population of the (1,1) and (2,2) energy levels of
\nhhh. The contours are as in Fig.~\ref{fg3}.
The (0,0) map position is as in Fig.~\ref{fg1}.
}
\label{fg10}
\end{figure*}

The coherence also leads to the nearly thermal \nhhh\ linewidths, 
which do not vary much in the interior of ammonia maps
(and which are usually narrower than the linewidths of the
surrounding C$^{18}$O and $^{13}$CO gas), 
and transonic to subsonic velocity dispersions (e.g., Smith \etal\ 2009).
We explore these issues in Fig.~\ref{fg10}.
Panel ({\bf a}) shows a dominant linewidth $\Delta v \sim 0.2$ \kms, which is very close to 
the thermal broadening of \nhhh\ at a kinetic temperature \Tkin\ = 10~K:
$\Delta v_{\rm th} \approx 0.16$ \kms\ (Eq.~(A.6) in L13).
The hyperfine structure fit to the \nhhh(1,1) multiplet
yields a maximum optical depth of $\sim 1$ for the isolated hfs component
$F'_1 \to F', F' \to F = 0\to 1, 1/2 \to 3/2$ at the offset (120\arcsec,$-40$\arcsec),
suggesting negligible ($<20$\%) optical depth broadening. 
Thus, the non-thermal motions do not dominate the thermal motions along the main axis E$\to$W of the filament.

However, approximately five times larger linewidths $\Delta v \sim 1$ \kms\
were detected at a few points spread over the northern and southern ridges
of the filament and in the vicinity of the outlier at the offset ($-200\arcsec, -40\arcsec$)
with a maximum $\Delta v = 1.0(1)$ \kms\ at $(\Delta\alpha,\Delta\delta) = (-240\arcsec, -40\arcsec)$.
The increasing contribution of the non-thermal motions to the \nhhh\ linewidths
at the filament bounds correlates with the dynamics of the surrounding 
C$^{18}$O and $^{13}$CO gas, showing
$\Delta v({\scriptscriptstyle {\rm C}^{18}{\rm O}}) = 1.3$ \kms\
and $\Delta v({\scriptscriptstyle ^{13}{\rm CO}}) = 1.9$ \kms\ at a
mean excitation temperature of 10~K (S94). 
At this temperature, the thermal width of the carbon monoxide molecules is 
$\Delta v_{\rm th} \sim 0.12$ \kms, i.e., non-thermal motions such as infall, 
outflow, or turbulence dominate over the thermal motions in the outskirts of the filament.
Therefore, the internal (C$^{18}$O) and external ($^{13}$CO) gas envelopes have a dynamical behavior that
differs from that of the central ammonia string.  

We also note that there are no visible correlations between the 
distribution of the \nhhh\ linewidths on the ammonia map in Fig.~\ref{fg10}{\bf a} and 
the directions of the jet and bipolar nebula outflows that are shown in Fig.~\ref{fg5}.
This can be understood if the axes of both jet and bipolar nebula were almost
orthogonal to the line of sight. 

Another way of looking at velocity coherence is to consider 
distributions of the Mach number defined locally as
\begin{equation}
 M_s = \sigma_{\rm turb}/c_s\ ,
\label{Eq5}
\end{equation}
where $\sigma_{\rm turb}$ is
the non-thermal velocity dispersion derived from the
apparent line width (Eq.~(A.7) in L13),
and $c_s$ is the thermal sound speed
\begin{equation}
c_s = 0.06\sqrt{T_{\rm kin}}\ ({\rm km}~{\rm s}^{-1})\  
\label{Eq6}
\end{equation}
for an isothermal gas (Eq.~(A.8) in L13).

Figure~\ref{fg10}{\bf b} shows the distribution of the calculated Mach numbers
at the 54 positions where \Tkin\ was measured directly from 
the relative population of the \nhhh(1,1) and (2,2) energy levels. 
Panel (\ref{fg10}{\bf b})  demonstrates that the ratio of
the non-thermal velocity dispersion to the sound speed along the main axis E$\to$W
lies in the interval $0.3 \la M_s \la 0.8$; i.e., 
the filament has subsonic internal velocity dispersion
in the coherent velocity region embedded in a generally
supersonic turbulent envelope characterized by
$M_s({\scriptscriptstyle {\rm C}^{18}{\rm O}}) \sim 3$ and
$M_s({\scriptscriptstyle ^{13}{\rm C}{\rm O}}) \sim 4$. 

Finally, we conclude that the filament consists of at least three 
kinematically distinguished fragments
with respect to their motion around the horizontal E$\to$W axis 
($\alpha\beta$, bar, and $\gamma$ fragments),
while all of them are involved in a common rotation around the vertical axis S$\to$N.
The $\alpha\beta$ and $\gamma$ fragments spin around the E$\to$W axis in opposite
directions that can be ascribed to dextral and sinistral chiralities. The bar shows no revolution, 
whereas the motion of the whole ammonia filament around the S$\to$N axis resembles
a simple solid-body rotation. All these properties imply that the filament is kinematically detached 
from the outer layer traced by $^{13}$CO emission.
The revealed coherence in the spatial and velocity distributions 
may, as usually supposed, be the manifestation of the intermittency of turbulent dissipation in molecular
clouds connected with processes of star formation (e.g., Hennebelle \& Falgarone 2012). 
The opposite chirality of the substructures in \object{L1251A} 
puts it in a list of fascinating objects  that show a filamentary structure of a complex helix-like geometry
(Carlqvist 2005).

\subsection{Mass of the filament}
\label{sect-3-3}

The mass of \nhhh\ gas can be estimated from the dynamical characteristics 
and linear scale of the filament.
If the cloud rotates as a solid body, 
the total mass of the ammonia filament stems from
equilibrium of the centrifugal and gravitational forces: 
\begin{equation}
{\cal M}_{\rm tot} = (\Omega^{\scriptscriptstyle\rm SN}_{\rm tot})^2 R^3/G\ , 
\label{Eq7}
\end{equation}
where $G$ is the gravitational constant, and $R$ is the radius 
along the major axis at which
the linear velocity $V = \Omega R$ is measured.
For $\Omega^{\scriptscriptstyle\rm SN}_{\rm tot} = 7\times10^{-15}$ \rds\ and
$R \approx 1.65$ pc, one finds 
${\cal M}_{\rm tot} \approx 45{\cal M}_\odot$. 
A two times higher mass was derived from the C$^{18}$O data in S94 for cores A and B: 
${\cal M}_{\scriptscriptstyle \rm A} \approx 56{\cal M}_\odot$, 
and ${\cal M}_{\scriptscriptstyle \rm B} \approx 38{\cal M}_\odot$.

If the considered object is an ellipsoid of revolution with axes
$\ell_1, \ell_2, \ell_3$, then its volume is ${\cal V} = \frac{\pi}{6}\ell_1\ell_2\ell_3$,
and for a uniform medium of mean density $\langle \rho\rangle$, its mass 
is ${\cal M} = {\cal V}{\langle \rho\rangle}$.
Substituting numerical values for $\ell_1, \ell_2, \ell_3$
and for ${\cal M} = {\cal M}_{\rm tot}$,
and using a mean molecular weight $\mu =2.33$,
we find the mean gas number density 
${\langle n\rangle}_{\scriptscriptstyle\rm H} = 
{\langle \rho\rangle}/\mu m_{\scriptscriptstyle\rm H} = 5.5\times10^3$ \cmm,
where $m_{\scriptscriptstyle\rm H}$ is the mass of the hydrogen atom.
A similar estimate,
${\langle n\rangle} = 6.0\times10^3$ \cmm,
is obtained if we adopt the rod-like geometry introduced in Sect.~\ref{sect-3-1}.

These gas densities are almost an order of magnitude lower than the 
${\langle n\rangle}_{\scriptscriptstyle \rm H_2} \sim 4\times10^4$ \cmm\
derived from ammonia emission along the major axis~--- 
a fact suggesting that there should be a gas number density gradient along the minor axes. 
Usually, the shape of a filament radial profile is described by 
a Plummer-like function of the form:
\begin{equation}
\rho(r) = {\rho_c}/{\left[ 1 + (r/r_0)^2 \right]^{\kappa/2}}\ ,
\label{Eq8}
\end{equation}
where $\rho_c$ is the central density of the filament, and $r_0$ is the radius
in the plane orthogonal to the major axis (Andr\'e \etal\ 2014).

For an isothermal gas cylinder in hydrostatic equilibrium, the power-law exponent
of the density profile is $\kappa = 4$ and
$n(r) \propto r^{-4}$ (Ostriker 1964).
However, observed filaments often show $n(r) \propto r^{-2}$, i.e., $\kappa \approx 2$
(e.g., Lada \etal\ 1999), suggesting that
dense filaments may not be strictly isothermal but better
described by a polytropic equation of state, 
$P(r) \propto \rho(r)^\gamma$ or $T(r) \propto \rho(r)^{\gamma-1}$ with $\gamma \la 1$
(Palmeirim \etal\ 2013). 
 
It is easy to show that our case also requires $\kappa \approx 2$ 
to match the aforementioned estimates of the gas number densities
$\langle n\rangle_{\scriptscriptstyle\rm H}$ and  ${\langle n\rangle}_{\scriptscriptstyle \rm H_2}$.
Thus, the filament \object{L1251A} is centrally condensed along its major axis,  
the \nhhh\ substructure is the densest part of it,
its density profile is close to the Plummer model with $\kappa \approx 2$, 
and the total mass of the ammonia substructure is about $45{\cal M}_\odot$.

\subsection{Gravitational equilibrium}
\label{sect-3-4}

The bound prestellar filaments are usually gravitationally unstable 
and prone to forming fragments (e.g., Andr\'e \etal\ 2014).
A self-gravitating cylinder will be in hydrostatic equilibrium 
only when its mass per unit length, 
\begin{equation}
{\cal M}_\ell = \int^{r_0}_0 2\pi r \rho(r) dr \ , 
\label{Eq9}
\end{equation}
has the special critical value (Ostriker 1964):
\begin{equation}
{\cal M}_{\ell, c} = \frac{2c^2_s}{G} = \frac{2kT}{G\mu m_{\scriptscriptstyle {\rm H}}} \approx
16.5T_{10}{\cal M}_\odot {\rm pc}^{-1}\ ,
\label{Eq10}
\end{equation}
where $c_s$ is the isothermal sound speed, and $T_{10}$ the gas temperature in units of 10~K.
Since the mean molecular weight $\mu = 2.33$ is expected to be almost
constant in the interstellar molecular clouds, the critical line mass ${\cal M}_{\ell, c}$
is independent of the gas density and determined only by the kinetic temperature \Tkin.

If the mass per unit length ${\cal M}_{\ell}$ is equal to ${\cal M}_{\ell, c}$, then
the self-gravitational force per unit mass and 
the pressure gradient force per unit mass are equal\footnote{A self-gravitating 
fluid usually assumes that thermal pressure is the
dominant source of gas pressure.}.
If ${\cal M}_{\ell} < {\cal M}_{\ell, c}$, then a filament expands until 
it is supported by external pressure.
Otherwise, when ${\cal M}_\ell > {\cal M}_{\ell, c}$, 
the filamentary structure is gravitationally unstable, and 
self-gravity dominates the pressure force. In this case,
if ${\cal M}_\ell \gg {\cal M}_{\ell, c}$, perturbations do not grow much
and the filament collapses toward the major axis
without fragmentation 
(Inutsuka \& Miyama 1992; Inutsuka \& Miyama 1997).

Adopting ${\langle T\rangle}_{\rm kin} \approx 10$~K (Sect.~\ref{sect-3-1}), 
one finds a critical mass per unit length for the ammonia filament of
${\cal M}_{\ell, c} \approx 17{\cal M}_\odot {\rm pc}^{-1}$.
Taking into account that the volumes of condensations ${\cal V}_{\alpha\beta} \approx {\cal V}_{\gamma}$,
their mean gas densities
${\langle n\rangle}_{\alpha\beta} \approx 2 {\langle n\rangle}_{\gamma}$,
and the projected lengths $\ell_{\alpha\beta} \approx 2.2$ pc, $\ell_{\gamma} \approx 1.1$ pc
(Sect.~\ref{sect-3-1}), 
the masses of the $\alpha\beta$ and $\gamma$ condensations are equal to
${\cal M}_{\alpha\beta} \sim 30 M_\odot$ and ${\cal M}_{\gamma} \sim 15 M_\odot$.
This gives the line masses of 
${\cal M}_{\ell, \alpha\beta} = {\cal M}_{\ell, \gamma} \sim 14{\cal M}_\odot {\rm pc}^{-1}$. 
Thus, the masses per unit length are less than or comparable to ${\cal M}_{\ell, c}$
which is expected for gas in hydrostatic equilibrium.

\section{Discussion}
\label{sect-4}

From the analysis of the dynamical stability of dark cloud \object{L1251}, 
Sato \etal\ (1994) concluded that all five  C$^{18}$O cores
embedded in molecular gas (see Fig.~\ref{fg1}) are gravitationally stable.
The present observations in \nhhh\ inversion lines
revealed that two of them, the cores A and B,
have an elongated filament-like morphology and a complex velocity field. 
The filament exists for about $10^7$ yr and
exhibits two types of global motions: ($i$) it rotates as a whole
around the minor S--N axis, and ($ii$) its eastern and western fragments revolve around 
the major E--W axis in opposite directions of dextral and sinistral chirality.
By analogy with the $\alpha$-$\omega$ geodynamo mechanism with broken symmetry 
(see, e.g., Fig.~2 in Love 1999), we suggest that 
these kinds of motions in \object{L1251A} may be maintained by helical magnetic fields
threaded through the cloud. 

Dynamo is a mechanism that converts kinetic energy into electromagnetic energy.
The motion of an electrical conductor through a magnetic field induces electrical currents
that can generate secondary induced magnetic fields that are sustained as long as energy
is supplied. A self-exciting dynamo requires no external fields or currents to sustain the
dynamo, aside from a weak seed magnetic field to get started. The $\omega$-effect is a
conversion of poloidal field to toroidal that is caused by differential rotation (shear)
of toroidal flows on the cloud surface. The $\alpha$-effect is a regeneration of the poloidal field
from toroidal field due to upwelling poloidal flows possessing vorticity.

In this picture,
the central region of \object{L1251A} (the bar between the eastern and western fragments), 
which shows no revolution, is the place where the 
helical magnetic field wrapped around the filament changes polarity from
negative and positive. As a result we observe the eastern and western fragments 
revolving in opposite directions.

The physical parameters measured in \object{L1251A} allow us to evaluate
the efficiency of the $\alpha$-$\omega$ dynamo mechanism.
The importance of magnetic induction relative to magnetic diffusion is characterized by
the magnetic Reynolds number (e.g., Brandenburg \& Subramanian 2005):
\begin{equation}
      R_{\rm m} = \sigma_{\rm rms}/({\tilde \eta} {\tilde k})\ ,
\label{Eq11}
\end{equation}
where $\sigma_{\rm rms}$ is the typical rms velocity, 
${\tilde \eta}$ the resistivity (cm$^2$~s$^{-1}$ in cgs units), and
${\tilde k} = 2\pi/{\ell}$ is the wavenumber.
If, for numerical estimate, we take for ${\tilde k}$ a lower limit on the size of eddies 
close to the molecular mean free path in a dense core, 
$\ell = (\sqrt{2} n \sigma)^{-1} \sim 6\times10^{-9}$ pc 
(the effective cross section $\sigma \sim 10^{-15}$ cm$^2$,
the gas density $n \sim 4\times10^4$ \cmm), then ${\tilde k} \sim 4\times10^{-10}$ cm$^{-1}$.

In dark clouds the gas is mostly neutral with low kinetic temperatures.
In this case, the resistivity is given by (e.g., Balbus \& Terquem 2001):
\begin{equation}
{\tilde \eta} = 234 x^{-1}_{\rm e} T^{1/2}\ ,
\label{Eq13}
\end{equation}
where $x_{\rm e} = n_{\rm e}/n$ is the ionization fraction, and $n$ the number density of neutral
particles.

If gas is shielded well from the external incident radiation, the gas temperature
mainly comes from the heating by cosmic rays.
Then the value of temperature is determined by the balance between heating and cooling. 
If the only source of heating are the cosmic rays and the cooling
comes from the line radiation, then
a lower bound on the kinetic temperature is about 8~K
(Goldsmith \& Langer 1978).

The ionization fraction $x_{\rm e}$ at the ionization-recombination equilibrium is
approximately given by (e.g., Draine \etal\ 1983):
\begin{equation}
x_{\rm e} = (\zeta_{\scriptscriptstyle \rm CR}/{\tilde \beta} n)^{1/2}\ ,
\label{Eq14}
\end{equation}
where $\zeta_{\scriptscriptstyle \rm CR}$ is the ionization rate by cosmic rays, and 
${\tilde \beta} = 3\times10^{-6} T^{-1/2}$ cm$^3$~s$^{-1}$ is the dissociative recombination
rate. Adopting the midplane value of
$\zeta_{\scriptscriptstyle \rm CR} \approx 10^{-17}$ s$^{-1}$ (e.g., Sano \& Stone 2002)
and the measured mean temperature \Tkin $\approx 10$~K for the ionization rate due to cosmic rays, we find
$x_{\rm e} \sim 2\times10^{-8}$, 
so that ${\tilde \eta} \sim 4\times10^{10}$ cm$^2$~s$^{-1}$. 
Then for $\sigma_{\rm rms} \sim 0.1$ \kms, we estimate the  
magnetic Reynolds number $R_{\rm m} \ga 600$. 

To control bulk motions, magnetic and non-thermal energy densities should be comparable,
\begin{equation}
B^2/8\pi \sim \rho \sigma^2_{\rm rms}/2\ , 
\label{Eq15}
\end{equation}
where $\rho$ is the gas
density, and $\sigma_{\rm rms}$ the rms non-thermal (turbulent) velocity.  
With $\sigma_{\rm rms} \sim 0.1$ \kms\ and 
$\rho = \mu m_{\scriptscriptstyle \rm H} n_{\scriptscriptstyle \rm H}    
\sim 3\times10^{-19}$ g~cm$^{-3}$,
the mean equipartition strength of the magnetic field for the filament \object{L1251A}
is expected to be $\langle B \rangle \sim 20$ $\mu$G,
which is a typical value of $B$ measured in dark clouds (e.g., Crutcher 2012).

To amplify a dynamically significant magnetic field from a background 
initially weak seed field in a self-gravitating system,
the gas must persist for a number of free-fall times,
which is (e.g., McKee, \& Ostriker 2007)
\begin{equation}
t_{ff} = \left( \frac{3\pi}{32 G \langle \rho \rangle } \right)^{1/2}  
\sim 2\times10^5~{\rm yr} \ ,
\label{Eq16}
\end{equation}
for \object{L1251A}.
The value of $t_{ff}$, together with the estimated periods of 
the filament spinning ($t \sim 10^7$ yr) and the age of the supernova remnants ($t \sim 10^{6-7}$ yr),
indicates that there has been plenty of time for the magnetic field to build up in the filament.

Theoretically, it has been suggested that a number of processes induce and maintain magnetic fields
in dark clouds. Among them, mechanisms of a linear growth with time 
of the interstellar seed field from $B \sim 1$~$\mu$G to a typical $B \sim 20$~$\mu$G inside dark clouds
are associated with a winding-up of frozen-in fields. Their efficiency is
proportional to the number of rotations. Since the period for a complete rotation of the filament
\object{L1251A} is comparable to its lifetime, which should be near the age of the supernova
remnants, this type of mechanism can be excluded.

Other processes are based on
an exponential growth that is realized in a dynamo amplification due to
shear flows, turbulent, and Coriolis motions. 
The exponential growth can, in principle, strengthen the magnetic field  
on a timescale that resembles the filament rotation period. 
In case \object{L1251A}, different stages of the filament formation may be characterized
by different types of magnetic field growth. In the first stages, when supernovae shock waves 
hit the cloud and caused strong turbulence,
weak seed magnetic fields were exponentially amplified by the small-scale 
turbulent dynamo (e.g., Beresnyak \& Lazarian 2015).
At present, turbulent motions in the centrally condensed gas along the major axis are subsonic
so that the turbulent dynamo is less effective.
On the other hand, complex kinematics of \object{L1251A} and
differential rotation may cause the Coriolis forces 
that tend to organize gas motions and electric currents into large scale Taylor columns (Taylor 1923)
aligned with the rotation axis 
that, in turn, lead to induction of a magnetic field similar to the $\omega$-dynamo effect.
The Taylor columns are formed when the Coriolis force 
is far more significant than the inertia force. 
The ratio between these two forces is characterized by the Rossby number, which 
is a dimensionless ratio of the inertia force over the Coriolis force:
\begin{equation}
{\cal R} = v/L\Omega\ , 
\label{Eq17}
\end{equation}
where $v$ is the velocity of a parcel of fluid (a small amount of higher density fluid),
$L$ is its scale (diameter), and $\Omega$ is the angular rotation. 
If the filament rotates more quickly than the parcel moves through it (i.e., ${\cal R} < 1$), 
the gas flow is dragged across the filament at right angles to the spin axis 
and, thus, lifts and twists magnetic field lines, which is a similar process to 
the $\alpha$-dynamo effect. 
Both $\alpha$- and $\omega$-dynamo can maintain the strength of the magnetic
field close to the equilibrium value given by Eq.~(\ref{Eq15}).

\section{Conclusions}
\label{sect-5}

Three hundred positions toward  the C$^{18}$O cores A and B within the
dark cometary-shaped cloud \object{L1251}
were observed in the \nhhh(1,1) and (2,2) inversion lines with the Effelsberg 100-m telescope
at a spectral resolution of 0.045 \kms\ and a spatial resolution of 40\arcsec.
The main results are summarized as follows.
\begin{enumerate}
\item[1.]
For the first time, we detect in \object{L1251}
a long and narrow structure covering a $38' \times 3'$ angular range ($\sim 3.3$ pc $\times$ 0.3 pc)
in the E--W direction.
The integrated ammonia intensity distribution is not homogeneous but concentrated in three 
condensations ($\alpha, \beta$, and $\gamma$) which form a rod-like filament.
All of them are involved in a common rotation around the vertical S--N axis
with angular velocity 
$\Omega^{\scriptscriptstyle\rm SN}_{\rm tot} \approx 7\times10^{-15}$ \rds.
\item[2.] 
The condensations exhibit a complex dynamical behavior: 
combined $\alpha$ and $\beta$ fragments are counter-rotating 
around the major E--W axis with respect to the $^{13}$CO envelope of the filament,
whereas the $\gamma$ fragment co-rotates with this envelope. 
For both of them, the angular velocity is 
$| \Omega^{\scriptscriptstyle\rm EW} | \approx 2\times10^{-14}$ \rds.
The central part of the filament between these two kinematically
distinct regions does not show any rotation around the E--W axis. 
\item[3.]
The dextral and sinistral  chirality of the $\alpha\beta$ and $\gamma$ condensations
indicates the presence of magnetic field helicity of two types, negative and positive,
supported by dynamo action.
\item[4.]
An exclusive feature of the filament is extremely narrow ammonia lines observed at 
several ``quiet zones'' where the linewidths are 
$\Delta v \sim 0.1$ \kms, meaning that they reveal 
almost purely thermal broadening at kinetic temperatures as low as $\sim 8$~K.
\item[5.]
Along the major E--W axis,  
we observed both the \nhhh(1,1) and (2,2) transitions at 54 positions and found the mean
ammonia column density
${\langle N\rangle}_{\scriptscriptstyle \rm NH_3} =
(1.03\pm0.05)\times10^{15}$ cm$^{-2}$ 
and the gas number density 
${\langle n\rangle}_{\scriptscriptstyle \rm H_2} = (3.9\pm1.3)\times10^4$ cm$^{-3}$ 
under the assumption that the beam filling factor $\eta = 1$. 
The mean kinetic temperature is $\langle T_{\rm kin} \rangle = 10.12\pm0.08$~K.
\item[6.]
The central part of the filament is organized into a coherent structure
akin to those found in some other star-forming dense cores. 
\item[7.]
The  Mach numbers calculated at the 54 positions within the coherent velocity region range between
0.3 and 0.8, which means that
the filament shows subsonic internal velocity dispersion.
The coherent velocity region is embedded in a generally
supersonic turbulent envelope with Mach numbers $\sim 3-4$ as derived
from observations of
${\rm C}^{18}{\rm O}$ and $^{13}{\rm C}{\rm O}$ lines. 
\item[8.]
The filament \object{L1251A} 
is centrally condensed along the E--W axis,  
the \nhhh\ substructure is the densest part of it, 
the density profile is close to the Plummer model with the power-law exponent
$\kappa \approx 2$, 
and the total mass of the ammonia substructure is ${\cal M} \sim 45{\cal M}_\odot$.
\item[9.]
The line masses ${\cal M}_{\ell}$ of the $\alpha\beta$ and $\gamma$ condensations 
${\cal M}_{\ell, \alpha\beta} = {\cal M}_{\ell, \gamma} \sim 14{\cal M}_\odot {\rm pc}^{-1}$
are less than or comparable to the critical value of ${\cal M}_{\ell, c} \sim 17{\cal M}_\odot {\rm pc}^{-1}$
which is expected for gas in hydrostatic equilibrium.

\end{enumerate}

Helical magnetic fields aligned with the spin E--W axis play an important role in the evolution of the filament.
The magnetic Reynolds number $R_{\rm m} \ga 600$ means that magnetic induction dominates magnetic diffusion, 
a condition required for $\omega$-dynamos. 
The Rossby number ${\cal R} < 1$ indicates that the Taylor columns 
are dragged across the filament leading to the $\alpha$-dynamo effect.
The joint action of both the $\omega$- and $\alpha$-dynamo  mechanisms
can provide a large scale magnetic field of positive and negative helicity that, in turn,
results in the observed gas motions of opposite chirality.

\begin{acknowledgements}
We thank the staff of the Effelsberg 100-m telescope for their assistance in observations,
and we appreciate Vadim Urpin's comments on an early version of the manuscript.
We also thank our referee G\"osta Gahm for suggestions that led to improvements in the paper.
SAL is grateful for the kind hospitality of 
the Max-Planck-Institut f{\"u}r Radioastronomie and Hamburger Sternwarte 
where this work was prepared.  
This work was supported in part by the grant DFG
Sonderforschungsbereich SFB 676 Teilprojekt C4, and
by the RFBR grant No.~14-02-00241.
\end{acknowledgements}

\clearpage
%----------------Figure 1

\clearpage
%----------------Figure 2

\clearpage
%----------------Figure 3

\clearpage
%----------------Figure 4

\clearpage
%----------------Figure 5

\clearpage
%----------------Figure 6

\clearpage
%----------------Figure 7

\clearpage
%----------------Figure 8

\clearpage
%----------------Figure 9

\clearpage
%----------------Figure 10

\clearpage
%-------------------------- Table 1
\begin{table}[t!]
\centering
\caption{Peak intensities of \nhhh(1,1) line emission 
and positions of the narrowest lines toward \object{L1251} A and B.
}
\label{tbl-1}
\begin{tabular}{c r@{,}l c r@{.}l r@{.}l }
\hline
\hline
\noalign{\smallskip}
Peak & \multicolumn{2}{c}{Offset$^b$} & 
\multicolumn{1}{c}{\Tmb$^c$} & \multicolumn{2}{c}{$V_{\scriptscriptstyle\rm LSR}$} 
& \multicolumn{2}{c}{$\Delta v$} \\[-2pt]
\multicolumn{1}{c}{ID$^a$} & {$\Delta\alpha$} & {$\Delta\delta$} 
& \multicolumn{1}{c}{(K)} & \multicolumn{2}{c}{(\kms)} & \multicolumn{2}{c}{(\kms)} \\
& ({\arcsec})&({\arcsec})  &  &\multicolumn{2}{c}{}  & \multicolumn{2}{c}{(FWHP)} \\
\noalign{\smallskip}
\hline

\noalign{\smallskip}
$\alpha$  & $160$&$-40$  & 4.5(5) & $-4$&212(3) & 0&260(8) \\ 
$\beta$   & $-400$&$0$   & 3.7(4) & $-3$&922(4) & 0&34(1)  \\ 
$\gamma$  &$-1160$&$0$   & 3.3(3) & $-3$&819(4) & 0&30(1)  \\ [2pt]
          &$-720$&$0$    & 2.2(2) & $-3$&984(6) & 0&24(1)  \\ 
          &$-1640$&$0$   & 0.8(1) & $-3$&69(2)  & 0&14(3)  \\ 
          &$-1480$&$-40$ & 0.9(1) & $-3$&67(1)  & 0&10(2)  \\ 
          &$-1360$&$40$ & 0.7(1) & $-3$&82(1)  & 0&21(3)  \\ 
          &$-1200$&$80$ & 0.7(1) & $-4$&12(3)  & 0&20(4)  \\ 
          &$-1080$&$80$  & 0.4(1) & $-4$&20(2)  & 0&18(5)  \\ 
          &$-1080$&$40$  & 0.8(1) & $-4$&11(2)  & 0&20(3)  \\ 
          &$-880$&$-80$  & 0.4(1) & $-3$&77(3)  & 0&18(4)  \\ 
          &$-760$&$40$  & 0.7(1) & $-4$&03(1)  & 0&12(2)  \\ 
          &$-680$&$0$  & 1.4(1) & $-3$&999(5)  & 0&20(2)  \\ 
          &$-640$&$0$  & 1.3(1) & $-3$&999(3)  & 0&13(2)  \\ 
          &$-600$&$0$  & 2.3(2) & $-4$&002(5)  & 0&21(1)  \\ 
          &$-520$&$120$ & 0.5(1) & $-4$&19(3)  & 0&16(5)  \\ 
          &$-160$&$80$ & 0.5(1) & $-4$&26(2)  & 0&19(3)  \\ 
          &$160$&$80$ & 0.5(1) & $-4$&18(3)  & 0&17(4)  \\ 
          &$360$&$-40$ & 2.8(3) & $-4$&303(4)  & 0&21(1)  \\ 
          &$400$&$-40$ & 2.0(2) & $-4$&318(6)  & 0&21(1)  \\ 
          &$400$&$-80$ & 2.8(3) & $-4$&334(4)  & 0&22(1)  \\ 
          &$440$&$-40$ & 1.9(2) & $-4$&298(7)  & 0&21(2)  \\ 
          &$480$&$-40$ & 1.1(1) & $-4$&31(1)  & 0&19(2)  \\ 

\noalign{\smallskip}
\hline
\noalign{\smallskip}
\multicolumn{8}{l}{{\bf Notes.} $^a$Greek letters label the peaks of ammonia } \\
\multicolumn{8}{l}{emission indicated in Fig.~\ref{fg2}. $^b$The zero offset (0,0) is}\\
\multicolumn{8}{l}{R.A. = 22:31:02.3, Dec = 75:13:39 (J2000).}\\
\multicolumn{8}{l}{$^c$The numbers in parentheses correspond to a $1\sigma$}\\
\multicolumn{8}{l}{statistical error on the last digit. }
\end{tabular}
\end{table}

\end{document}